\documentclass[useAMS,usenatbib]{mn2e}
\usepackage{multirow}
\usepackage{graphicx}
\usepackage{amssymb} 
\usepackage{mn2e-breakabs}
\def\prl{Phys. Rev. Lett.}

\begin{document}

\title[Cosmological implications of the BOSS-CMASS $\xi(s)$]
{
The clustering of galaxies in the SDSS-III Baryon Oscillation
 Spectroscopic Survey: cosmological implications of the large-scale two-point correlation function
}
\author[A.G. S\'anchez et al.]
{\parbox[t]{\textwidth}{
Ariel~G. S\'anchez$^{1}$\thanks{E-mail: arielsan@mpe.mpg.de},
C.~G. Sc\'occola$^{2,3}$,
A.~J. Ross$^{4}$,
W. Percival$^{4}$,
M. Manera$^{4}$,
F. Montesano$^{1}$,
X. Mazzalay$^{1}$,
A.~J. Cuesta$^{5}$,
D.~J. Eisenstein$^{6}$,
E. Kazin$^{7}$, 
C.~K. McBride$^{6}$,
K. Mehta$^{8}$,
A.~D. Montero-Dorta$^{9}$,
N. Padmanabhan$^{5}$,
F. Prada$^{10,11,9}$,
J. A. Rubi\~no-Mart\'{\i}n$^{2,3}$,
R. Tojeiro$^{4}$,
X. Xu$^{8}$,
M. Vargas Maga\~na$^{12}$,
E. Aubourg$^{12}$,
N.~A. Bahcall$^{13}$,
S. Bailey$^{14}$,
D. Bizyaev$^{15}$,
A.~S. Bolton$^{16}$,
H. Brewington$^{15}$,
J. Brinkmann$^{15}$,
J.~R. Brownstein$^{16}$,
J. Richard Gott, III$^{13}$,
J.~C. Hamilton$^{12}$,
S. Ho$^{14,17}$,
K. Honscheid$^{18}$,
A. Labatie$^{12}$,
E. Malanushenko$^{15}$,
V. Malanushenko$^{15}$,
C. Maraston$^{4}$,
D. Muna$^{19}$,
R.~C. Nichol$^{4}$,
D. Oravetz$^{15}$,
K. Pan$^{15}$,
N.~P. Ross$^{14}$,
N.~A. Roe$^{14}$,
B. A. Reid$^{14,20}$,
D. J. Schlegel$^{14}$,
A. Shelden$^{16}$,
D.~P. Schneider$^{21,22}$,
A. Simmons$^{15}$,
R. Skibba$^{8}$,
S. Snedden$^{15}$,
D. Thomas$^{4}$,
J. Tinker$^{19}$,
D.~A. Wake$^{23}$,
B.~A. Weaver$^{19}$,
David H. Weinberg$^{24}$,
Martin White$^{25,14}$,
I. Zehavi$^{26}$,
and G. Zhao$^{4,27}$
}
\vspace*{6pt} \\ 
$^{1}$ Max-Planck-Institut f\"ur extraterrestrische Physik, Postfach 1312, Giessenbachstr., 85741 Garching, Germany.\\ 
$^{2}$ Instituto de Astrof{\'\i}sica de Canarias (IAC), C/V{\'\i}a L\'actea, s/n, La Laguna, Tenerife, Spain. \\
$^{3}$ Dpto. Astrof{\'\i}sica, Universidad de La Laguna (ULL), E-38206 La Laguna, Tenerife, Spain.\\
$^{4}$ Institute of Cosmology \& Gravitation, University of Portsmouth, Dennis Sciama Building, Portsmouth PO1 3FX, UK.\\
$^{5}$ Department of Physics, Yale University, 260 Whitney Ave, New Haven, CT 06520, USA.\\
$^{6}$ Harvard-Smithsonian Center for Astrophysics, 60 Garden St., Cambridge, MA 02138. \\
$^{7}$ Centre for Astrophysics and Supercomputing, Swinburne University of Technology, P.O. Box 218, Hawthorn, Victoria 3122, Australia.\\
$^{8}$ Steward Observatory, University of Arizona, 933 N. Cherry Ave., Tucson, AZ 85721, USA.\\
$^{9}$ Instituto de Astrofisica de Andalucia (CSIC), Glorieta de la Astronomia, E-18080 Granada, Spain.\\ 
$^{10}$ Campus of International Excellence UAM+CSIC, Cantoblanco, E-28049 Madrid, Spain.\\
$^{11}$ Instituto de Fisica Teorica (UAM/CSIC), Universidad Autonoma de Madrid, Cantoblanco, E-28049 Madrid, Spain.\\
$^{12}$ APC, University of Paris Diderot, CNRS/IN2P3, CEA/IRFU, Observatoire de Paris, Sorbonne Paris Cit\'e, France.\\  
$^{13}$ Department of Astrophysical Sciences, Princeton University, Peyton Hall, Princeton, NJ 08540, USA.\\
$^{14}$ Lawrence Berkeley National Laboratory, 1 Cyclotron Rd, Berkeley, CA 94720, USA.\\
$^{15}$ Apache Point Observatory, P.O. Box 59, Sunspot, NM 88349-0059, USA.\\
$^{16}$ Department of Physics and Astronomy, The University of Utah, 115 S 1400 E, Salt Lake City, UT 84112, USA\\
$^{17}$ Department of Physics, Carnegie Mellon University, 5000 Forbes Ave., Pittsburgh, PA 15213, USA.\\
$^{18}$ Department of Physics and CCAPP, Ohio State University, Columbus, OH, USA.\\
$^{19}$ Center for Cosmology and Particle Physics, New York University, NY 10003, USA.\\ 
$^{20}$ Hubble Fellow.\\
$^{21}$ Department of Physics and Astronomy, Pennsylvania State University, University Park, PA 16802, USA.\\
$^{22}$ Institute for Gravitation and the Cosmos, The Pennsylvania State University, University
Park, PA 16802, USA.\\
$^{23}$ Yale Center for Astronomy and Astrophysics, Yale University, New Haven, CT, USA.\\
$^{24}$ Department of Astronomy and CCAPP, Ohio State University, Columbus, OH, USA.\\
$^{25}$ Department of Physics, University of California Berkeley, CA 94720, USA.\\
$^{26}$ Department of Astronomy, Case Western Reserve University, Cleveland, OH 44106, USA.\\
$^{27}$ National Astronomy Observatories, Chinese Academy of Science, Beijing, 100012, P.R.China.\\
}
\date{Submitted to MNRAS}

\maketitle
\begin{abstract}
We obtain constraints on cosmological parameters from the spherically averaged
redshift-space correlation function of the CMASS Data Release 9 (DR9) sample of the
Baryonic Oscillation Spectroscopic Survey (BOSS).
We combine this information with additional data from recent CMB, SN and BAO measurements.
Our results show no significant evidence of deviations from the standard flat-$\Lambda$CDM model,
whose basic parameters can be specified by $\Omega_{\rm m} = 0.285\pm0.009$,
$100\,\Omega_{\rm b} = 4.59\pm0.09$, $n_{\rm s} = 0.961\pm 0.009$,
 $H_{0}=69.4\pm0.8 \, {\rm km} \,{\rm s}^{-1}\,{\rm Mpc}^{-1}$ and $\sigma_{8} = 0.80 \pm 0.02$.
The CMB+CMASS combination sets tight constraints on 
the curvature of the Universe, with $\Omega_{k}=-0.0043\pm0.0049$, and the tensor-to-scalar amplitude ratio,
for which we find $r<0.16$ at the 95 per cent confidence level (CL).
These data show a clear signature of a deviation from scale-invariance also in the presence of tensor modes, with $n_{\rm s}<1$ at
the 99.7 per cent CL. We derive constraints on the fraction of massive neutrinos of $f_{\nu}<0.049$ (95 per cent CL),
implying a limit of $\sum m_{\nu} < 0.51 {\rm eV}$.
We find no signature of a deviation from a cosmological constant from the combination of all datasets,
with a constraint of $w_{\rm DE}=-1.033\pm0.073$ when this parameter is assumed time-independent,
and no evidence of a departure from this value when it is allowed to evolve as $w_{\rm DE}(a)=w_0+w_a(1-a)$. 
The achieved accuracy on our cosmological constraints
is a clear demonstration of the constraining power of current cosmological observations.
\end{abstract}


\section{Introduction}
\label{sec:intro}
In recent years, a wealth of  precise cosmological observations have been used to place tight
constraints on the values of the fundamental cosmological parameters \citep[e.g.][]{Riess1998,Perlmutter1999,
Spergel2003,Riess2004,Tegmark2004,Sanchez2006,Spergel2007, Riess2009,Komatsu2009,Sanchez2009,Komatsu2010,
Percival2010,Reid2010,Riess2011,Blake2011,Montesano2012}.
The unexpected conclusion from these studies is that we seem to live in a more
complex and richer Universe than originally suspected; one which is currently
undergoing a phase of accelerating expansion.
Understanding the origin of cosmic acceleration is one of the most outstanding
problems in physics as it may hold the key to a true revolution in our understanding of the Universe.

Within the context of general relativity, cosmic acceleration implies that the
energy-density budget of the Universe is dominated by a dark energy component,
which counteracts the attractive force of gravity.
A key parameter that can be used to characterize this component is the dark energy equation
of state $w_{\rm DE}$, defined as the ratio of its pressure to density.
In the standard $\Lambda$CDM model, dark energy can be described by a fixed equation of state
specified by $w_{\rm DE} = -1$, which can be interpreted as the quantum energy of the vacuum.
However, a large variety of alternative models have been proposed, which
predict different values of $w_{\rm DE}$ and its possible evolution with time \citep[for a review see e.g.][]{Peebles2003,Frieman2008,Gott2011}.

Measurements of the large-scale structure (LSS) of the Universe are expected to play a major role at shedding light on the causes of cosmic acceleration.
The shape of the galaxy power spectrum, $P(k)$, and its Fourier transform, the two-point correlation
function $\xi(r)$, encode useful information which can be used to obtain robust constraints,
not only on dark energy, but also on other important physical parameters like neutrino masses,
the curvature of the Universe or details of inflationary physics \citep{Percival2002,Tegmark2004,Cole2005,Sanchez2006,Spergel2007,Komatsu2009,
Komatsu2010,Percival2010,Reid2010,Keisler2011,Blake2011,Montesano2012}.
A special feature of large-scale clustering measurements provides a powerful method to
probe the expansion history of the Universe: the baryon acoustic oscillations (BAO).
 These are a series of small amplitude oscillations imprinted on the power spectrum
\citep{Eisenstein1998,Meiksin1999}, which are
analogous to the acoustic oscillations present in the cosmic microwave background (CMB) power spectrum.
In the correlation function these are transformed into a single peak whose position is
related to the sound horizon at the drag redshift \citep{Matsubara2004}. As this scale can be calibrated to high precision
from CMB observations, BAO measurements at different redshifts can be used as a standard ruler to
measure the distance-redshift relation \citep{Blake2003,Linder2003}.
The BAO feature was first detected in the clustering pattern of the luminous red galaxy (LRG) sample of the
Sloan Digital Sky Survey \citep[SDSS,][]{York2000} by \citet{Eisenstein2005} and the Two-degree Field
Galaxy Redshift survey
\citep[2dFGRS,][]{Colless2001,Colless2003} by \citet{Cole2005} and has been subsequently observed using a variety of datasets and techniques \citep{Hutsi2006,Padmanabhan2007,Percival2007,Percival2010,Cabre2009,Gaztanaga2009,
Kazin2010,Beutler2011,Blake2011,Ho2012,Seo2012}.

Driven by the potential of LSS observations for shedding light on the problem of
the nature of dark energy, several ground-breaking galaxy surveys are currently being
constructed or designed which will be substantially larger than their predecessors.
The ongoing Baryonic Oscillation Spectroscopic Survey \citep*[BOSS,][]{Schlegel2009} is 
an example of these new surveys. BOSS is a part of SDSS-III \citep{Eisenstein2011} aimed at
obtaining redshifts for $1.5\times10^6$ massive galaxies out to $z=0.7$ over an area of $10,000\,{\rm deg}^2$.
This information will provide a high-precision determination of the expansion history of
the Universe through accurate measurements of the BAO feature in the large-scale galaxy clustering.
BOSS will also attempt to obtain, for the first time, BAO measurements at high redshift ($z \approx 2.5$)
through the Ly$\alpha$ forest absorption spectra of about 150,000 quasars.

\begin{figure*}
\centerline{\includegraphics[width=0.9\textwidth]{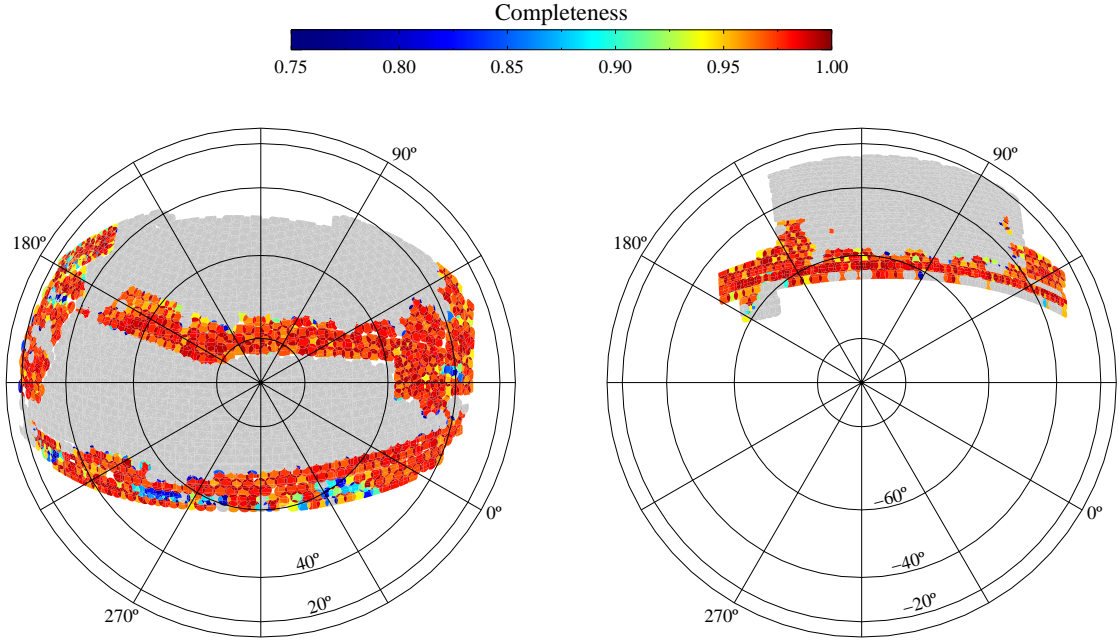}}
\caption{
The sky coverage, in Galactic coordinates, of the CMASS DR9 spectroscopic sample used in this analysis
in the northern (left panel) and southern
(right panel) Galactic hemispheres. Different sectors are colour-coded according to their completeness.
The low completeness at many edges is due to planned but currently unobserved tiles that will
overlap the current geometry.
The light grey shaded region shows the expected footprint of the final survey, totalling 10,269 deg$^2$.
}
\label{fig:map}
\end{figure*}

The increasing precision of the new surveys demands accurate models of the
LSS observations to extract the maximum amount of information from the data without introducing
biases or systematic effects. The BAO signal in the correlation function and power spectrum is modified
by the non-linear evolution of density fluctuations, redshift-space distortions, and galaxy bias
\citep{Meiksin1999,Eisenstein2007, Seo2007,Seo2008,Smith2008, Angulo2008,Crocce2008,Sanchez2008,Gott2009,Kim2009,Montesano2010,Kim2011}.
These effects must be taken into account in the models used to interpret the observations.
New developments in perturbation theory, such as Renormalized Perturbation Theory \citep[RPT, ][]{Crocce2006}, have provided substantial progress regarding the theoretical understanding of the
effects of non-linear evolution, which can now be accurately modelled
\citep{Crocce2006, Matsubara2008, Matsubara2008b, Taruya2009}, and even partially corrected for
\citep{Eisenstein2007,Seo2010,Padmanabhan2012}.
Based on RPT, \citet{Crocce2008} proposed a simple model to describe the full shape
of the correlation function on large scales. \citet{Sanchez2008} showed that this model yields
an excellent description of the results of N-body simulations, providing a robust tool to extract unbiased
cosmological constraints out of measurements of $\xi(r)$.
\citet{Sanchez2009} used this model to obtain constraints on cosmological parameters from the correlation
function of a sample of LRGs from SDSS-DR6 \citep{Adelman-McCarthy2008}
as measured by \citet{Cabre2009}.
The same ansatz has been used by \citet{Beutler2011} and \citet{Blake2011} for the analysis of the
correlation functions of the 6dF and WiggleZ galaxy surveys, respectively. 
An analogous approach was used by
\citet{Montesano2012} to study the cosmological implications of the LRG power spectrum in SDSS-DR7 \citep{Abazajian2009}.

In this paper we apply the parametrization of \citet{Crocce2008} to the redshift-space
correlation function of a high redshift galaxy sample from BOSS Data Release 9 (DR9).
This sample, denoted CMASS, is constructed through a set of colour-magnitude cuts designed to
select a roughly volume-limited sample of massive, luminous galaxies 
\citep[][Padmanabhan et al. in prep.]{Eisenstein2011}.
We combine the CMASS clustering information with recent measurements
of CMB, BAO and type Ia supernovae data. We derive constraints on the parameters of the standard
$\Lambda$CDM model, and on a number of potential extensions, with an emphasis on
the constraints on the dark energy equation of state.
Our analysis is part of a series of papers aimed at providing a thorough and comprehensive
description of the galaxy clustering in the CMASS sample 
\citep{Blanton2012,Manera2012,Reid2012,Ross2012,Samushia2012,Tojeiro2012}.

The outline of this paper is as follows. In Section~\ref{sec:corfunc} we describe the galaxy sample
that we use and the procedure we follow to compute its correlation function. We also present a discussion
on the cosmological information contained in this measurement. Section~\ref{sec:moredata} describes the
additional datasets that we combine with the CMASS correlation function to obtain constraints on cosmological parameters.
Our model of the full shape of the correlation function, the parameter spaces we explore and the applied methodology 
is described in Section~\ref{sec:method}. In Section~\ref{sec:results} we present
our results for constraints on cosmological parameters from different combinations of datasets and parameter
spaces. In Section~\ref{ssec:northsouth} we analyse the differences in the clustering of the northern and southern Galactic hemispheres and
explore their implications on the obtained cosmological constraints. Finally,
Section~\ref{sec:conclusions} contains our main conclusions.

\section{Clustering analysis of the BOSS-CMASS galaxies}
\label{sec:corfunc}

We base our analysis on the large-scale two-point correlation function, $\xi(s)$, of the BOSS-CMASS galaxy sample.
In this Section we review the most important details of the construction of 
the sample (Section~\ref{ssec:cmass}), and our clustering analysis (Section~\ref{ssec:clustering}).

\subsection{The CMASS galaxy sample}
\label{ssec:cmass}

The galaxy target selection of BOSS consists of two separate samples, dubbed LOWZ and CMASS, designed to cover different redshift ranges
\citep[][Padmanabhan et al. in prep.]{Eisenstein2011}. These samples are selected on the basis of photometric observations done with the dedicated 2.5-m Sloan
Telescope \citep{Gunn2006}, located at Apache Point Observatory in New Mexico, using a drift-scanning mosaic
CCD camera \citep{Gunn1998}. These samples are constructed on the basis of $gri$ colour cuts designed to select luminous galaxies at different
redshifts at a roughly constant number density. Spectra of the LOWZ and CMASS samples are obtained using
the double-armed BOSS spectrographs, which are significantly upgraded from those used by SDSS-I/II,
covering the wavelength range 3600~\AA{} to 10000~\AA{} with a resolving power of 1500 to 2600 \citep{Smee2012}.
Spectroscopic redshifts are then measured using the minimum-$\chi^2$ template-fitting procedure described in
\citet{Aihara2011}, with templates and methods updated for BOSS data as described in \citet{Bolton2012}.

Our analysis is based on the clustering properties of the CMASS sample,
which is selected to be an approximately complete galaxy sample down to a limiting stellar
mass \citep{Maraston2012}. The CMASS sample is dominated by early type galaxies, although it contains a significant
fraction of massive spirals \citep[$\sim$26 per cent,][]{Masters2011}.
Most of the galaxies in this sample are central galaxies, with a $\sim$10 per cent satellite fraction \citep{White2011,Nuza2012}.

\citet{Blanton2012} presents a detailed description of the construction of the catalogue for LSS studies
based on this sample, and the calculation of the completeness of each sector of the survey mask, that is,
the areas of the sky covered by a unique set of spectroscopic tiles \citep[][]{Blanton2003},
which we characterize using the {\sc Mangle} software \citep[][]{Hamilton2004, Swanson2008}.
We only include sectors with completeness larger than 75\%. Our results are not affected by this
limit, as this leaves out only a small fraction of the total DR9 area. 
We restrict our analysis to the redshift range $0.43 < z < 0.7$, producing a final sample
of 262,104 galaxies, of which 205,947 and 56,157 are located in the Northern
and Southern Galactic hemispheres, respectively. 
Fig.~\ref{fig:map} shows the angular footprint, in Galactic coordinates, of the resulting sample
for the Northern (left) and Southern (right) Galactic caps (hereafter NGC and SGC, respectively),
colour-coded according to sector completeness. 

\citet{Nuza2012} compared the small and intermediate-scale clustering of this sample 
to the expectations of a flat $\Lambda$CDM cosmological model by applying an abundance matching
technique to the Multidark simulation.
In three companion papers, \citet{Reid2012}, \citet{Samushia2012} and \citet{Tojeiro2012} 
study the signature of redshift-space distortions in this sample and explore its cosmological implications.
Here we focus on the shape of the large-scale monopole correlation function to obtain constraints on
cosmological parameters.

\subsection{The redshift-space correlation function }
\label{ssec:clustering}

We characterize the clustering of the CMASS galaxy sample by means of the angle-averaged
redshift-space two-point correlation function $\xi(s)$. Here we summarize the 
procedure we follow to obtain this measurement.

The first step in the calculation of three-dimensional clustering statistics is the conversion of the observed
redshifts into distances. For this we assume a flat $\Lambda$CDM fiducial cosmology
with matter density, in units of the critical density, of $\Omega_{\rm m}=0.274$, and a Hubble parameter
$h=0.7$ (expressed in units of $100\,{\rm km}\,{\rm s}^{-1}{\rm Mpc}^{-1}$). This is the same fiducial 
cosmology assumed by \citet{White2011} and our companion papers \citep{Blanton2012,Manera2012,Ross2012,Reid2012,Tojeiro2012}.
As will be discussed in Section~\ref{ssec:info}, the choice of the fiducial cosmology has implications on the 
resulting correlation function.

We then compute the full correlation function $\xi(s,\mu)$, where $\mu\equiv s_{||}/|\vec{s}|$
and $s_{||}$ is the radial component of the separation vector $\vec{s}$,
using the estimator of \citet{Landy1993}, namely
\begin{equation}
\xi(s,\mu) = \frac{DD- 2 DR + RR}{ RR},
\label{eq:lys}
\end{equation}
where $DD$, $DR$ and $RR$ are the normalized pair counts in each bin of $(s,\mu)$ in the data
and a random sample with 50 times more objects than the original data, constructed to follow
the same selection function \citep[for more details on the construction of the random catalogue see][]{Blanton2012}.
We infer the angle-averaged redshift-space correlation function as the monopole of $\xi(s,\mu)$, that is
\begin{equation}
\xi(s) = \frac{1}{2}\int_{-1}^{1} \xi(s,\mu) {\rm d}\mu.
\end{equation}
This method should be preferred over the commonly used one, in which the $DD$, $DR$, and $RR$ counts are
integrated over $\mu$ before they are combined as in equation~(\ref{eq:lys}) to compute $\xi(s)$,
ignoring the fact that the geometry of the survey
introduces a $\mu$ dependence on $RR$ \citep{Samushia2011,Kazin2012}, although the differences between the
two approaches are more significant for higher multipoles.

 When computing the pair counts in equation~(\ref{eq:lys}), a few important corrections must be taken into
account. This is done by assigning a series of weights to each object in the real and random catalogues.
First, we apply a radial weight given by 
\begin{equation}
w_{\rm r}=\frac{1}{1+P_{w}\bar{n}(z)},
\label{eq:wradial}
\end{equation}
where $\bar{n}(z)$ is the expected number density of the catalogue at the given redshift and $P_{w}$ is a
free parameter. \citet{Hamilton1993} showed that setting $P_{w}=4\pi J_3(s)$,
where $J_3(s)=\int_0^s\xi(s')s'^2{\rm d}s'$, minimizes the
variance on the measured correlation function for the given scale $s$. 
Following standard practice we use a scale-independent value of $P_w=2\times 10^4\,h^{-3}{\rm Mpc}^3$.
\citet{Reid2012} show that the full scale-dependent weight provides only a marginal improvement over the 
results obtained using this constant value.

We include additional weights to account for non-random contributions to the sample incompleteness and
to correct for systematic effects. The incompleteness in a given sector of the mask has a random component
due to the fact that not all galaxies satisfying the CMASS selection criteria are observed spectroscopically.
In any clustering measurement this is taken into account by down-sampling the random catalogue in that region
of the sky by the same fraction.
However, there are two other sources of missing redshifts which require special treatment: redshift
failures and fibre collisions.

Even when the spectrum of a galaxy is observed, it might not be possible to obtain a
reliable estimation of the redshift of the object, leading to what is called a redshift failure.
As shown in \citet{Ross2012}, 
the probability that a spectroscopic observation leads to a redshift failure is
not uniform across the field since these tend to happen for fibres
 located near the edges of the observed plates. 
 Hence, these missing redshifts  cannot be considered as
an extra component affecting the overall completeness of the sector.

However, the main cause of missing redshift is fibre collisions
\citep[][]{Zehavi2002, Masjedi2006}. The BOSS spectrographs are fed by
optical fibres plugged on plates, which must be separated by at least 62$''$. It is then not possible
to obtain spectra of all galaxies with neighbours
closer than this angular distance in one single observation. The problem is alleviated in
sectors covered by multiple exposures but, in general, it is impossible to observe all the
objects in crowded regions.

\begin{figure}
\centering
\centerline{\includegraphics[width=\columnwidth]{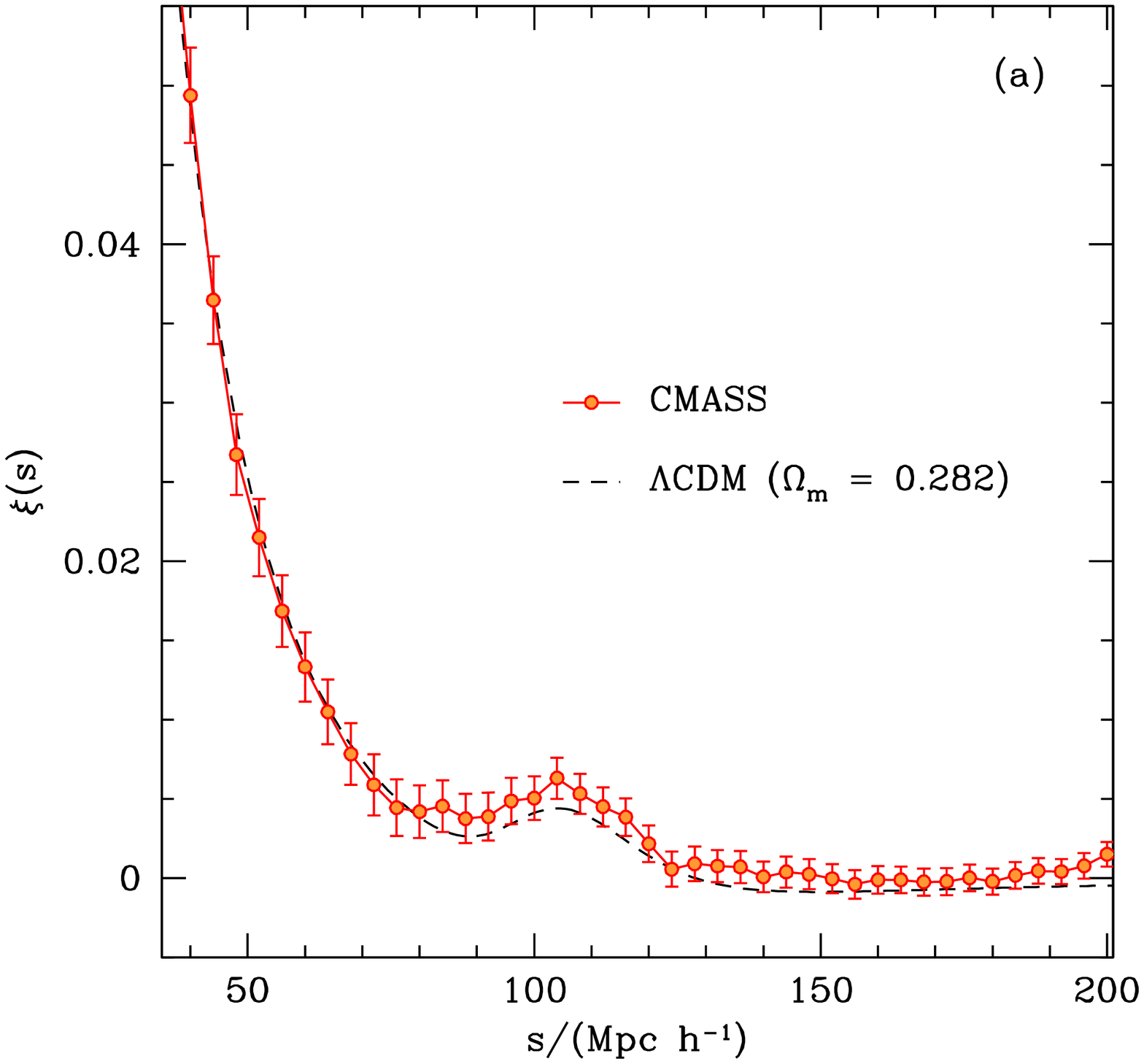}}
\centerline{\includegraphics[width=\columnwidth]{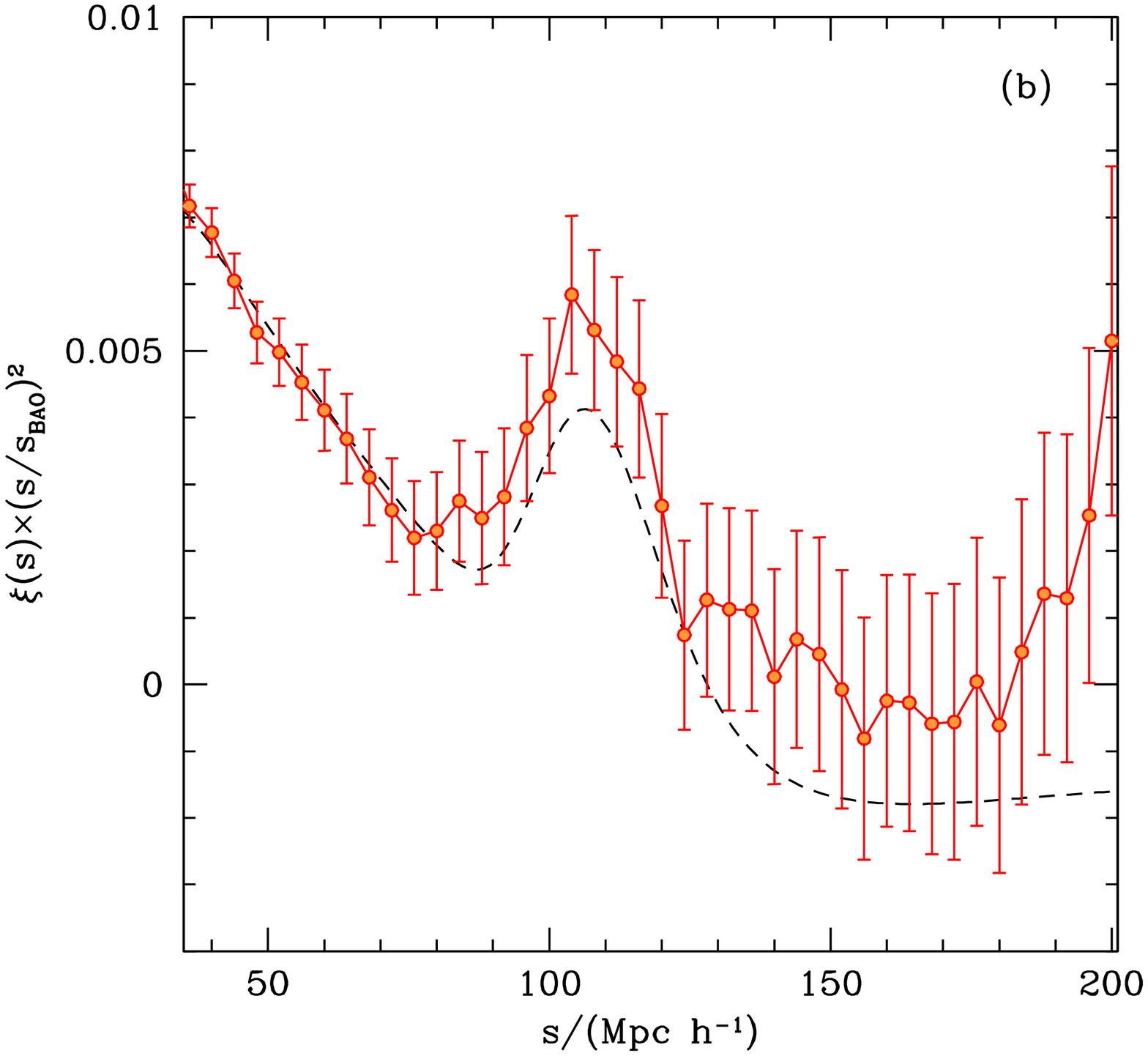}}
\caption{
Panel (a): spherically averaged redshift-space two-point correlation function of the full CMASS
sample. The errorbars were obtained from a set of 600 mock catalogues constructed 
to follow the same selection function of the survey \citep{Manera2012}. The dashed line corresponds
to the best-fitting $\Lambda$CDM model obtained by combining the information from the shape of the
correlation function and CMB measurements (see Section \ref{ssec:lcdm}). Panel (b): same as panel 
(a), but rescaled by $(s/s_{\rm BAO})^2$, where
$s_{\rm BAO}=153.2 \, {\rm Mpc}$ (which corresponds to 107.2 $h^{-1}{\rm Mpc}$),
to highlight the baryonic acoustic feature.
}
\label{fig:corfunc}
\end{figure}

To correct for these effects we follow \citet{Ross2012} and implement two sets of weights,
$w_{\rm rf}$ and $w_{\rm fc}$, whose default value is $1$ for all
galaxies in the sample. For every galaxy with a redshift failure, we
increase by one the value of $w_{\rm rf}$ of the nearest galaxy with a
good redshift measurement.  Similarly, for each galaxy whose
redshift was not observed due to fibre collisions, the value of $w_{\rm fc}$ of
its neighbour, closer than 62$''$, is increased by one. 
These are then combined into a single weight to correct for missing redshifts given by
\begin{equation}
 w_{\rm mr} = w_{\rm rf}+w_{\rm fc}-1.
\label{eq:wangular}
\end{equation}
On the scales analysed in this paper, the application of these weights
effectively corrects for the effects of fibre collisions and redshift failures
providing an excellent agreement with the results obtained using the method
recently proposed by \citet{Guo2011}.
 
\citet{Ross2012} performed a detailed analysis of the systematic effects that could potentially
affect any clustering measurement based on the CMASS sample showing that,
besides redshift failures and fibre collisions, other important
corrections must be considered in order to obtain unbiased clustering measurements.
They found that the local stellar density is the dominant source of systematic errors as it has a significant
effect on the probability of detecting a CMASS galaxy. In this way, the variations of stellar density across the
sky introduce spurious fluctuations in the galaxy density field which affect all clustering measurements.
\citet{Ross2012} found that this systematic effect can be corrected for by applying a set of weights
$w_{\rm sys}$ which depend on both the stellar density and the galaxy $i_{\rm fiber2}$ magnitude,
that is, the $i$-band magnitude measured within a 2$''$ aperture.
We include these weights in the final total weight $w_{\rm tot}$ used
in all our clustering measurements
\begin{equation}
 w_{\rm tot} = w_{\rm r}\,w_{\rm mr}\,w_{\rm sys}.
\label{eq:wtotal}
\end{equation}

Additional potential systematics such as Galactic extinction,
seeing, airmass, and sky background have also been investigated,
and all have been found to potentially introduce much smaller spurious fluctuations.
These non-cosmological fluctuations can be corrected for using a weighting
scheme that minimises these fluctuations as a function of a
given systematic effect.

The upper panel of Fig.~\ref{fig:corfunc} shows the large-scale redshift-space correlation function
of the full CMASS sample obtained through the procedure described above. 
The dashed line corresponds to the best-fitting $\Lambda$CDM model obtained from the combination of 
this measurement with CMB observations as described in Section~\ref{ssec:lcdm}.
The BAO peak can be seen more clearly in the lower panel,
which shows the same measurement rescaled by the ratio $(s/s_{\rm BAO})^2$, where $s_{\rm BAO}=153.2\,{\rm Mpc}$
corresponds to the sound horizon scale in our fiducial cosmology. 
As will be discussed in more detail in Section~\ref{ssec:northsouth}, the measurements of the two-point
correlation function in the NGC and SGC sub-samples exhibit intriguing differences. Although the overall
shapes of these measurements are similar, they show differences at the scale of the acoustic peak.
In Section~\ref{ssec:northsouth} we discuss the significance of these differences
and their impact on the inferred cosmological constraints.

To obtain an estimate of the covariance matrix of the correlation function measured in these samples,
we use a set of $N_{\rm m}=600$ independent mock catalogues based on a method similar to {\sc PTHalos} 
\citep{Scoccimarro2002}, although with some important differences. A detailed description of the construction of
these mock catalogues and a comparison with the results of N-body simulations is presented in 
\citet{Manera2012}\footnote{These mock catalogues will be made
available in http://www.marcmanera.net/mocks/}.
These simulations correspond to the same fiducial cosmology used to 
measure $\xi(s)$ in the real catalogue and were designed to follow the selection function of 
the NGC and SGC CMASS sub-samples. We measured the correlation function of each mock catalogue using the same
binning schemes as for the real data and the radial weights of equation~(\ref{eq:wradial}).
We then use these measurements to obtain an estimate of the covariance matrix of $\xi(s)$ 
in the NGC, SGC as
\begin{equation}
C_{ij} =  \frac{1}{(N_{\rm m}-1)}
\sum_{\rm k=1}^{N_{\rm m}} \left(\xi_{\rm k}(s_{\rm i})-{\bar\xi}(s_{\rm i})\right)
\left(\xi_{\rm k}(s_{\rm j})-{\bar\xi}(s_{\rm j})\right),
\label{eq:covmat} 
\end{equation}
where $\xi_{\rm k}(s)$ is the correlation function from the $k$-th mock catalogue, and ${\bar\xi}(s)$ is the 
mean correlation correlation function from the ensemble of realizations.
As in \citet{Ross2012}, we assume that the NGC and SGC regions are independent and compute the covariance
matrix of the full CMASS sample as
$\mathbfss{ C}_{\rm full}^{-1}=\mathbfss{ C}_{\rm NGC}^{-1}+\mathbfss{ C}_{\rm SGC}^{-1}$.
The errorbars in Fig.~\ref{fig:corfunc} correspond to the square root of the diagonal entries in
$\mathbfss{ C}_{\rm full}$. 

\section{Additional data-sets}
\label{sec:moredata}

As described in Section~\ref{ssec:info}, the two-point correlation function contains valuable
cosmological information. However, it is not possible to constrain high-dimensional parameter spaces to high 
precision using this measurement alone. Here we describe the additional datasets with which we combine 
the CMASS $\xi(s)$ in order to improve the obtained cosmological constraints. 

Undoubtedly, the measurements of the temperature and polarization fluctuations of the CMB constitute the
most powerful and robust cosmological probe to date. In particular, the results from the seven-year of observations of the WMAP
satellite \citep{Hinshaw2008} and the South Pole Telescope \citep[SPT,][]{Keisler2011}
provide a detailed picture of the structure of the acoustic peaks in the CMB power spectrum up to 
multipoles $\ell\simeq3000$. This information places tight constraints on the
parameters of the basic $\Lambda$CDM model. However, the power of these observations is limited by nearly exact degeneracies that arise when deviations
from this simple model are explored \citep{Efstathiou1999}.
These degeneracies can be broken by combining the CMB information with additional datasets, such as the shape of $\xi(s)$.

In our analysis we use the temperature power spectrum in the range $2 \leq \ell \leq 1000$ and the
temperature-polarization power spectrum for $2 \leq \ell \leq 450$ from the seven-year of observations of the WMAP
satellite \citep{Jarosik2010,Komatsu2011,Larson2011},
combined with the recent SPT observations of \citet{Keisler2011} for $650 \leq \ell \leq 3000$. While for $\ell \lesssim 650$ the CMB power
spectrum is dominated by primary anisotropies, at smaller angular scales it contains a non-negligible contribution from secondary anisotropies.
To take this into account, we follow the treatment of \citet{Keisler2011} and include the contributions from the Sunyaev-Zel'dovich (SZ) effect,
and the emission from foreground galaxies (considering both a clustered and a Poisson point source contribution)
in the form of templates whose amplitudes are considered as nuisance parameters and marginalized over.
These templates are only applied to the SPT data.
We refer to the WMAP-SPT combination as our ``CMB'' dataset.

Additionally, we consider the constraints provided by the Hubble diagram of 
type Ia supernovae (SN) obtained from the compilation of \citet{Conley2011}.
This sample contains 472 SN, combining the high-redshift SN from the first three years of the 
Supernova Legacy Survey (SNLS) with other samples, primarily at lower redshifts.
In order to take into account the effect of the systematic errors in our cosmological constraints
we follow the recipe of \citet{Conley2011}, who performed a detailed analysis of all identified systematic uncertainties,
characterizing them in terms of a covariance matrix that incorporates effects such as
the recently discovered correlations between SN luminosity and host galaxy properties,
as well as the uncertainties of the empirical light-curve models.
When only SN data are used to constrain cosmological parameters, the uncertainty budget
is dominated by statistical errors. However, when these data are combined with external datasets, as in our case,
statistical and systematic uncertainties are comparable, highlighting the importance of an accurate
treatment of the later.

We also use information from other clustering measurements in the form of 
constraints on $y_s(z)$ and $A(z)$ from independent BAO analyses. 
We use the results of \citet{Beutler2011}, who
obtained an estimate of $y_{\rm s}(z=0.106)= 0.336 \pm 0.015$
from the large-scale correlation function of the 6dF Galaxy Survey \citep[6DFGS,][]{Jones2009}.
We also include the 2\% distance measurement of $(y_{\rm s}(0.35))^{-1}= 8.88\pm0.17$ recently obtained by 
\citet{Padmanabhan2012} and \citet{Xu2012} from the application of an updated version of the
reconstruction technique
proposed by \citet{Eisenstein2007} to the clustering of galaxies from the final SDSS-II LRG sample.
The application of this algorithm resulted in an improvement of almost a factor two in the accuracy 
on $y_{\rm s}$ over the constraint obtained from the unreconstructed sample.
 We combine the result from these analyses into our ``BAO'' dataset.
In a recent analysis, \citet{Blake2011} used the full shape of the two-point correlation function
from the final dataset of the WiggleZ Dark Energy Survey \citep{Drinkwater2010} to obtain constraints
on $y_{\rm s}(z)$ and $A(z)$ for three independent redshift slices of width $\Delta z = 0.4$.
We do not include these measurements in our analysis given the significant  overlap 
of the WiggleZ data with the sample analysed here. However, as shown in \citet{Blanton2012},
the WiggleZ BAO measurements are in excellent agreement with those inferred from the
CMASS sample.

The datasets described above are used in different combinations to check the 
consistency of the constraints returned. We start from the constraints obtained using CMB
data alone, which we then combine with the CMASS correlation function in our ``CMB+CMASS''
combination. We then add separately the SN and additional BAO data to test the impact of these datasets
on the obtained results. Our tightest constraints are obtained from the combination of all
four datasets.

\section{Methodology}
\label{sec:method}

We obtain constraints on cosmological parameters following a similar approach to that of
\citet{Sanchez2009}. In this Section we summarize the main points of our analysis method.
The parametric model we use to describe the shape of the correlation function in
redshift space is summarized in Section~\ref{ssec:rpt}. 
Section~\ref{ssec:param} describes the different parameter sets that we consider,
together with the methodology we follow to explore and constrain them.
Section~\ref{ssec:info} describes the way in which cosmological information is extracted out of 
a measurement of $\xi(s)$.

\subsection{Modelling the full-shape of $\xi(s)$}
\label{ssec:rpt}

Following \citet{Crocce2008} and \citet{Sanchez2008}, we model the shape of the large-scale correlation function,
$\xi(s)$, by applying the following parametrization:
\begin{equation}
 \xi(s) = b^2 \left[\xi_{\rm L}(s)\otimes {\rm e}^{-(k_{\star}s)^2} 
+ A_{\rm MC} \,\xi'_{\rm L}(s)\,\xi^{(1)}_{\rm L}(s) \right], 
\label{eq:xi_model}
\end{equation}
where $b$, $k_{\star}$ and $A_{\rm MC}$ are treated as free parameters, and the symbol $\otimes$ 
denotes a convolution. Here $\xi'_{\rm L}$ is the derivative of the linear correlation 
function $\xi_{\rm L}$, and $\xi^{(1)}_{\rm L}$ is defined by 
\begin{equation}
 \xi_{\rm L}^{(1)}(s) \equiv \hat{s} \cdot \nabla^{-1}\xi_{\rm L}(s)
=\frac{1}{2\pi^2}\int P_{\rm L}(k)\,j_1(ks)k\,{\rm d}k ,
\label{eq:xi1}
\end{equation}
with $j_{\rm 1}(y)$ denoting the spherical Bessel function of order one.
This parametrization  was originally proposed by \citet{Crocce2008} and
it is based on the theoretical framework of 
Renormalized Perturbation Theory \citep[RPT,][]{Crocce2006}, where
the non-linear power spectrum $P_{\rm NL}(k,z)$ can be computed as the sum of two terms
\begin{equation}
 P_{\rm NL}(k, z) = G(k,z)^2P_{\rm L}(k,z) + P_{\rm MC}(k,z).
\label{eq:pk_rpt}
\end{equation}
The first of these contributions represents a re-summation in the renormalized propagator, $G(k, z)$,
of all the terms in the perturbation theory expansion of $P_{\rm NL}(k,z)$ proportional
to the linear theory power spectrum $P_{\rm L}(k)$. 
The second term groups all the remaining contributions, which arise from the coupling of different Fourier modes. 
The non-linear correlation function is then given by an analogous decomposition, which motivates
the parametrization of equation~(\ref{eq:xi_model}). The exponential in the first term of equation~(\ref{eq:xi_model})
is based on the fact that, in the high-$k$ limit, the propagator can be accurately described as a Gaussian
damping, while the second term corresponds to the leading order contribution to $\xi_{\rm MC}$
arising from the coupling of two initial modes.

\citet{Sanchez2008} compared this model against the results of an ensemble of large volume N-body simulations
\citep[L-BASICC-II,][]{Angulo2008} at various redshifts, and showed that it provides an accurate description
of the full shape of the correlation function, including also the effects of bias and redshift-space distortions. 
\citet{Sanchez2009} used this model to obtain constraints on cosmological parameters from the correlation function
of the LRG sample from SDSS-DR6 measured by \citet{Cabre2009}.
This parametrization has also been used by \citet{Beutler2011} and \citet{Blake2011} for their analyses of the
correlation function measurements from the 6dF Galaxy Survey and the WiggleZ Dark Energy Survey.
\citet{Montesano2012} applied an analogous parametrization to study the cosmological implications of the
power spectrum of an LRG sample drawn from SDSS-DR7.

The smoothing length $k_{\star}$ depends on cosmology and redshift, but also on galaxy
type through its dependence on halo mass. For this reason, we follow a conservative approach and consider
$k_{\star}$ as a free parameter.

Following \citet{Sanchez2009}, we restrict the comparison of the model of equation~(\ref{eq:xi_model}) and
the measured CMASS correlation function to $s > 40\,h^{-1}{\rm Mpc}$. Although this is a conservative lower
limit, on smaller scales further contributions to $\xi_{\rm MC}(s)$ should be considered. We also limit 
our analysis to scales $s<200\,h^{-1}{\rm Mpc}$, since on larger scales all viable models predict 
similar shapes for $\xi(s)$.
We compute the likelihood of the model assuming a Gaussian form ${\cal L}\propto\exp(-\chi^2/2)$.
This choice is justified by the results of \citet{Manera2012}, who found that the probability distribution
function of $\xi(s)$ inferred from the ensemble of mock catalogues can be described by a Gaussian distribution
to high accuracy.

To allow for the fact that, when computing the CMASS correlation function, galaxy distances were calculated with
our fiducial cosmology, a correction must be applied to the model before computing its 
corresponding $\chi^2$ value (see Section~\ref{ssec:info}).

\subsection{Cosmological parameter spaces}
\label{ssec:param}

The starting point of our analysis is the basic $\Lambda$CDM parameter
space. This is the simplest model able to successfully describe a large variety
of cosmological datasets. 
It corresponds to a flat universe where the energy budget contains contributions from 
cold dark matter (CDM), baryons, and dark energy, which is given 
by vacuum energy or a cosmological constant $\Lambda$ (i.e. with an equation of state parameter
$w_{\rm DE} = -1$). Primordial density fluctuations are adiabatic, Gaussian, and have
a power-law spectrum of Fourier amplitudes, with a negligible contribution from tensor modes.
This model can then be defined by specifying the values of the
following six parameters:
\begin{equation}
{\bf P}_{\rm \Lambda CDM} = (\omega_{\rm b}, \omega_{\rm dm}, \Theta, \tau,A_s,n_s). 
\label{eq:param}
\end{equation}
The baryon and dark matter densities, $\omega_{\rm b} = \Omega_{\rm b}h^2$ and 
$\omega_{\rm dm} = \Omega_{\rm dm}h^2$,  
and the ratio between the horizon scale at recombination
and the angular diameter distance to the corresponding redshift, $\Theta$, characterize
the homogeneous background model. This set is equivalent to fixing the values of $\Omega_{\rm b}$,
$\Omega_{\rm dm}$ and $h$, but it is better constrained by the CMB data.
The primordial power spectrum of the scalar fluctuations is described by
its amplitude, $A_{\rm s}$, and spectral index, $n_{\rm s}$.
The values of these parameters are quoted at the pivot wavenumber of $k= 0.05\,{\rm Mpc}^{-1}$.
Finally, $\tau$ gives the optical depth to the last scattering surface, which we compute assuming
instantaneous reionisation. Our constraints on the $\Lambda$CDM parameter space are described in 
Section~\ref{ssec:lcdm}.

In order to constrain possible deviations from the $\Lambda$CDM model, in 
Sections~\ref{ssec:omk}--\ref{ssec:darkenergy} we explore a number of
extensions of this parameter space by allowing for variations on the following set
of parameters:
\begin{equation}
{\bf P}_{\rm extra} = (\Omega_{\rm k}, f_{\nu},  r, w_{\rm DE}). 
\label{eq:paramextra}
\end{equation}
These are the curvature of the Universe, the dark matter fraction in the form of massive neutrinos, $f_{\nu}=\Omega_{\nu}/\Omega_{\rm dm}$,
the tensor-to-scalar mode amplitude ratio of the primordial fluctuations\footnote{When including tensor modes
we assume the slow-roll consistency relation and fix the tensor spectral index as $n_{\rm t}=-r/8$.}, and the
dark energy equation of state parameter. For most of this paper, we assume that the dark energy equation of state 
is independent of redshift. In Section \ref{ssec:wa} we allow also for a time variation
of this parameter using the standard linear parametrization of \citet{Chevallier2001} and \citet{Linder2003}
given by
\begin{equation}
w_{\rm DE}(a) = w_0 + w_a(1-a), 
\label{eq:wa}
\end{equation}
where $a$ is the expansion factor and $w_0$ and $w_a$ are the parameters we constrain.

We also present constraints on other quantities which can be derived from the ones listed in
equations~(\ref{eq:param}) and (\ref{eq:paramextra}).
In particular we are interested in:
\begin{equation}
{\bf P}_{\rm der} = (\Omega_{\rm DE}, \Omega_{\rm m}, h, \sigma_8, t_0, z_{\rm re}, D_{\rm V}(z_{\rm m}),\sum m_{\nu}, f).
\label{eq:paramder}
\end{equation}
These are the dark energy and total matter densities (i.e., including the contributions from baryons, cold dark matter and neutrinos),
the Hubble factor, the rms linear perturbation theory variance in spheres of radius $8\,h^{-1}{\rm Mpc}$, the age of the universe,
the redshift of reionization, the average distance to the mean redshift of the sample (given by equation~\ref{eq:dv}),
the sum of the neutrino masses, given by
\begin{equation}
\sum m_{\nu}= 94.4\,\omega_{\rm dm}f_{\nu}\, {\rm eV}, 
\label{eq:mnu}
\end{equation}
 and the logarithmic derivative of the growth factor, $f(z_m)={\rm d}\ln D/{\rm d}\ln a$.  

We explore these parameter spaces using the {\sc CosmoMC} code of \citet{Lewis2002}.
{\sc CosmoMC} uses {\sc camb} to compute power spectra for 
the CMB and matter fluctuations \citep{Lewis2000}.
We use a generalized version of {\sc camb} which supports a time-dependent 
dark energy equation of state \citep{Fang2008}.
We included additional modifications from \citet{Keisler2011} and \citet{Conley2011} to compute
the likelihood of the SPT and SNLS datasets. 

In order to compare a given cosmological model with the datasets described in
Sections~\ref{sec:corfunc} and \ref{sec:moredata} it is necessary to
include a set of nuisance parameters given by
\begin{equation}
\mathbf{P_{\rm nuisance} }\equiv (b,k_{\star},A_{\rm MC},D^{\rm SZ}_{3000},D^{\rm PS}_{3000},D^{\rm CL}_{3000},\alpha,\beta),
\end{equation}
to the parameter sets described above.
The bias factor $b$, the damping scale $k_{\star}$ and the mode-coupling amplitude $A_{\rm MC}$
are described in detail in Section~\ref{ssec:rpt}. The quantities $D^{\rm SZ}_{3000}$, $D^{\rm CL}_{3000}$ and 
$D^{\rm PS}_{3000}$
give the amplitudes of the contributions from the Sunyaev-Zel'dovich effect, the clustering of the foreground
emissive galaxies and 
their shot-noise fluctuation power, respectively, to the high-$\ell$ CMB angular power spectrum. 
The foreground terms are used only when calculating the SPT likelihood; they are not used when calculating the WMAP likelihood.
We follow \citet{Keisler2011} and apply Gaussian priors on the amplitude of each of these foreground terms given
by $D^{\rm PS}_{3000} = 19.3 \pm 3.5\,\mu{\rm K}^2$, $D^{\rm CL}_{3000} = 5.0 \pm 2.5\,\mu{\rm K}^2$, and 
$D^{\rm SZ}_{3000} = 5.5 \pm 3.0\,\mu{\rm K}^2$.
The parameters $\alpha$ and $\beta$ are additional nuisance parameters introduced by \citet{Conley2011} for the
correct treatment of the systematics in the analysis of the SN data.
When quoting constraints on the parameters of equations~(\ref{eq:param})-(\ref{eq:paramextra}), 
the values of these parameters are marginalized over.

\begin{figure}
\centering
\centerline{\includegraphics[width=\columnwidth]{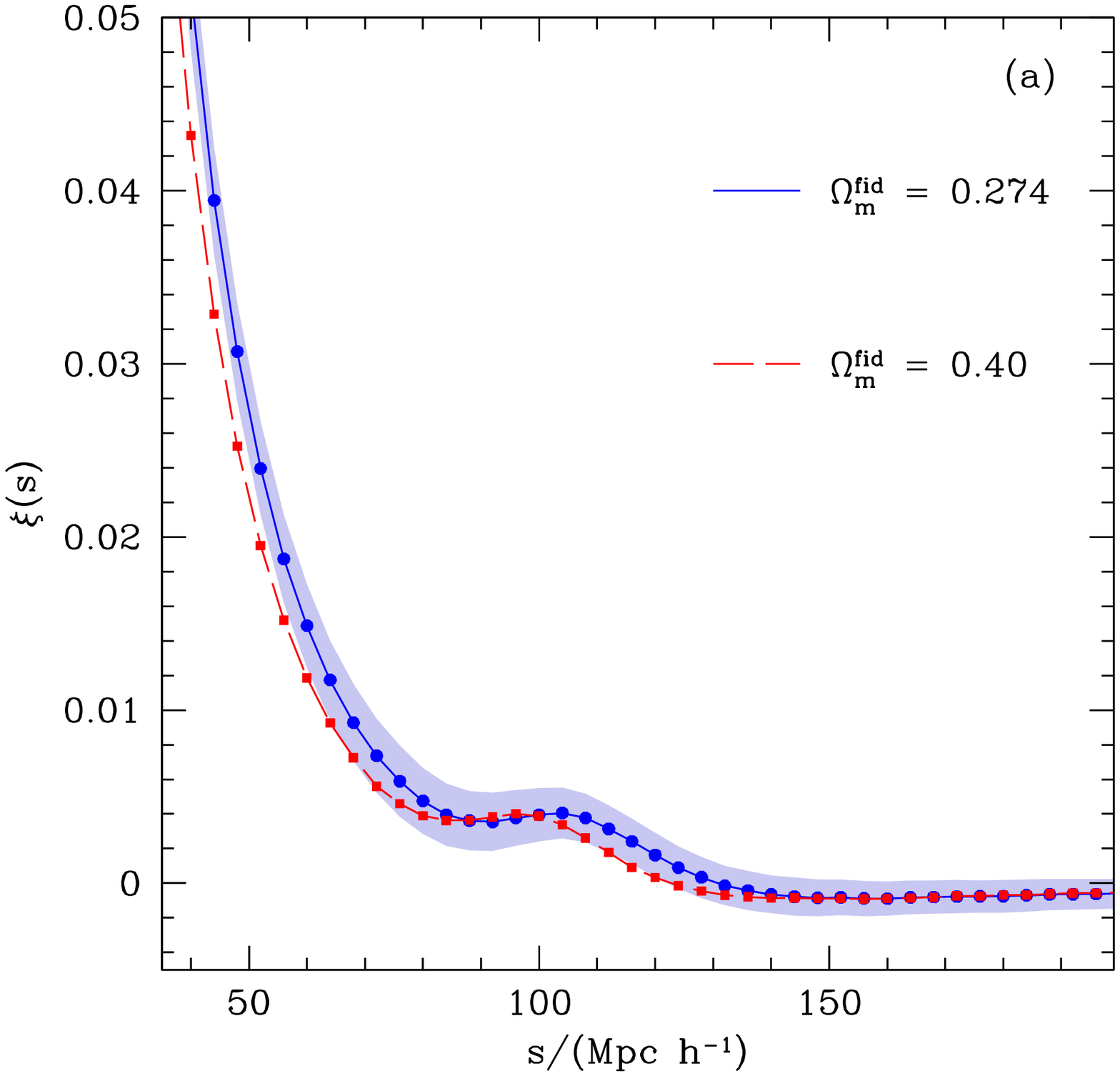}}
\centerline{\includegraphics[width=\columnwidth]{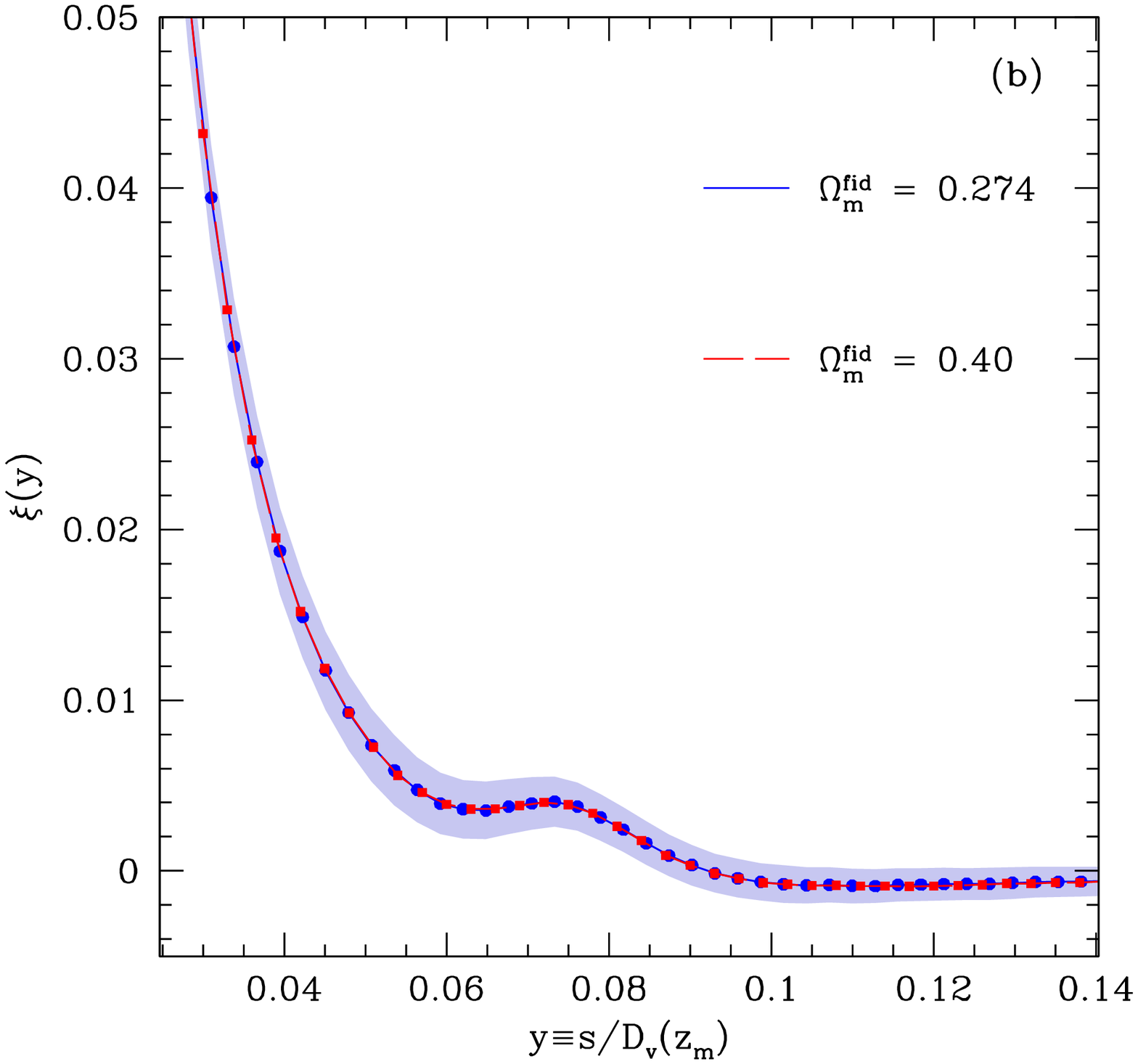}}
\caption{
Panel (a): mean correlation function from our ensemble of mock catalogues obtained by assuming the true cosmological parameters as fiducial values
(circles connected by a solid line) and a flat cosmology with $\Omega_{\rm m}=0.4$ (squares connected by a dashed line). The shaded region correspond to the 
variance between the different realizations of the ensemble. Panel (b): same measurements as panel (a),
but expressed as a function of $y=s/D_{\rm V}(z_{\rm m})$, which removes 
the dependence on the fiducial cosmology.
}
\label{fig:fiducial}
\end{figure}

\subsection{Extracting information out of $\xi(s)$}
\label{ssec:info}

In this Section we describe the information encoded in the shape of the two-point correlation function and 
how it can be used to obtain constraints on cosmological parameters.
As described in Section~\ref{ssec:clustering}, the measurement of the correlation function requires the assumption of
a fiducial cosmology to map the observed redshifts into distances. This fact has important implications 
on the parameter combinations that are constrained by $\xi(s)$.

 Different choices of the fiducial cosmology lead to a rescaling of the distances to the individual
galaxies $s\rightarrow s'$, affecting the volume element of the survey. 
This effect can be encapsulated in the Jacobian of the transformation \citep{Eisenstein2005,Sanchez2009,Kazin2012}
\begin{equation}
 {\rm d}^3s'=\left(\frac{D_{\rm V}'(z_{\rm m})}{D_{\rm V}(z_{\rm m})}\right)^3{\rm d}^3s.
\label{eq:jacobian}
\end{equation}
Here $D_{\rm V}(z_{\rm m})$ is a measure of the average distance to the mean redshift of the survey, $z_{\rm m}=0.57$,
given by
\begin{equation}
 D_{\rm V}(z)=\left((1+z)^2D_{\rm A}(z)^2\frac{cz}{H(z)}\right)^{1/3}
\label{eq:dv}
\end{equation}
where $H(z)$ is the Hubble parameter and $D_{\rm A}(z)$ is the proper angular diameter distance.

Fig.~\ref{fig:fiducial} illustrates the effect of assuming different fiducial cosmologies on
the measurement of $\xi(s)$. The points connected by a solid line in panel (a) show the mean correlation function of
our ensemble of mock catalogues, obtained assuming as fiducial cosmology the true values of the simulation parameters.
The shaded region corresponds to the variance between the individual realizations. The squares connected by a dashed
line correspond to  the mean correlation function from the same set of mock catalogues, but obtained assuming a flat $\Lambda$CDM model with
$\Omega_{\rm m}=0.4$. The two measurements show significantly different slopes and positions of the acoustic peak. 
As equation~(\ref{eq:jacobian}) suggests, this change is simply due to a rescaling of the horizontal axis.
This effect can be better appreciated in panel (b) of Fig.~\ref{fig:fiducial}, where the impact of the fiducial cosmology 
has been removed by expressing the measured correlation functions in terms of the dimensionless variable
$y\equiv s/D_{\rm V}^{\rm fid}(z_{\rm m})$. This exercise shows that, although the true underlying correlation
function is not a real observable, it is possible to obtain a measurement which is independent of the fiducial
cosmology by expressing it as $\xi(y)$. 

The particular choice of the fiducial cosmology must be taken into account when comparing a measurement of $\xi(s)$ with
theoretical predictions. As described above, this can be achieved by expressing both model and measurements
in terms of $y$. Alternatively, the 
effect of the fiducial cosmology might be introduced in the model by rescaling the scales $s$ by a factor
\begin{equation}
\gamma = \frac{D_{{\rm V}}^{\rm fiducial}(z_{\rm m})}{D_{\rm V}^{{\rm model}}(z_{\rm m})},
\label{eq:cor_fact}
\end{equation}
before comparing it to the measured $\xi(s)$. We follow this approach in our analysis.

\begin{figure}
\centering
\centerline{\includegraphics[width=\columnwidth]{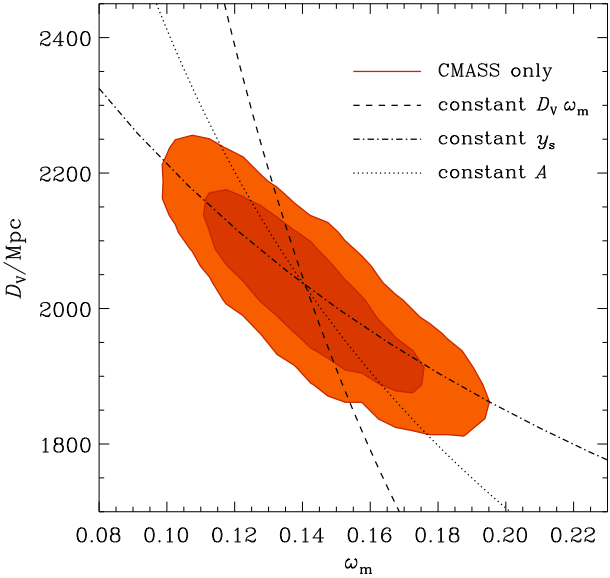}}
\caption{ 
The 68 and 95 per cent marginalized constraints in the $\omega_{\rm m}-D_{\rm V}(z_{\rm m})$ plane, where 
$\omega_{\rm m}\equiv \Omega_{\rm m} h^2$, obtained from the shape of the CMASS correlation function alone (solid lines).
The dashed, dot-dashed and dotted lines correspond to constant values of $D_{\rm v}(z_{\rm m})\,\omega_{\rm m}$, 
$y_{\rm s}(z_{\rm m})$ (equation ~\ref{eq:ys}), and $A(z_{\rm m})$ (equation \ref{eq:apar}), respectively.
}
\label{fig:om_dv}
\end{figure}

 The most important source of cosmological information in $\xi(s)$ is the location of the acoustic peak, which is
closely related to the sound horizon at the drag redshift $r_{\rm s}(z_{\rm d})$. 
Associating the position of the peak in $\xi(y)$ with this scale, it is clear that this measurement will 
provide constraints on the parameter combination
\begin{equation}
y_{\rm s}(z_{\rm m})=\frac{r_{\rm s}(z_{\rm d})}{D_{\rm V}(z_{\rm m})}.
\label{eq:ys}
\end{equation}
However, the location of the acoustic peak does not correspond exactly to the acoustic scale. Non-linear evolution
and redshift space distortions damp the acoustic peak and shift its position towards smaller scales 
\citep{Smith2008,Crocce2008,Angulo2008,Sanchez2008}. Nonetheless, if these effects are modelled correctly,
a measurement of $\xi(s)$ would still provide constraints on the parameter combination of equation~(\ref{eq:ys}),
allowing for the correct underlying cosmology to be recovered.

 Similarly, the measurement of the power spectrum, $P(k)$, will be subject to the same effect,
which can be removed by multiplying the measured wave-numbers by $D_{\rm V}(z_{\rm m})$.
In this way, the wavelength of the acoustic oscillations inferred from 
a measurement of $P(k)$ will provide constraints on $y_{\rm s}^{-1}(z_{\rm m})$. As $P(k)$ is not a dimensionless quantity, its amplitude is
also affected by the fiducial cosmology (by a factor proportional to $(D_{\rm V}(z_{\rm m}))^3$).
This can be avoided by working with the dimensionless power spectrum $\Delta^2(k)=P(k)k^3/(2\pi^2)$.

Besides the BAO, the power spectrum contains information on another useful scale.
The location of the turn-over in $P(k)$ is related to the size of the sound horizon at the time of
matter-radiation equality. In the absence of massive neutrinos, and for a fixed effective number 
of relativistic species, this scale is $k_{\rm eq}\propto \Omega_{\rm m}h^2\,{\rm Mpc}^{-1}$. 
Taking into account the effect of the fiducial cosmology, the quantity that can actually be constrained is 
$k_{\rm eq}D_{\rm v}(z_{\rm m})$.
The information about this parameter combination
is also encoded in the shape of the correlation function, where it is
related to the position of the zero-crossing at scales larger than those of the acoustic peak
\citep{Prada2011}. 
In this way, a measurement of $\xi(s)$ provides constraints on the same parameter combination.
This quantity is degenerate with other parameters, like the baryon density and the scalar
spectral index, which also affect the shape of $\xi(s)$.
However, the later are tightly constrained by CMB observations
\citep[e.g.,][]{Komatsu2011,Keisler2011}. 

The contours in Fig.~\ref{fig:om_dv} show the two-dimensional marginalized constraints in
the $\omega_{\rm m}-D_{\rm V}(z_{\rm m})$ plane, where  $\omega_{\rm m}\equiv \Omega_{\rm m} h^2$, 
obtained from the shape of the CMASS correlation function, using the model described in Section~\ref{ssec:rpt}.
To ameliorate the effect of the degeneracies between $\omega_{\rm m}$, and $\omega_{\rm b}$ and $n_{\rm s}$
in this exercise, we have applied Gaussian priors of $\omega_{\rm b}=0.0222\pm0.0010$
and $n_{\rm s}=0.966\pm0.020$. These priors are weaker than the corresponding accuracy with which these parameters are 
determined by current CMB data (see Section~\ref{sec:results}), allowing us to quantify more clearly the
information provided by $\xi(s)$. The full combination of this measurement with CMB data will result in
slightly tighter constraints. 

The dashed and dotted lines in Fig.~\ref{fig:om_dv} correspond to constant values of
$y_{\rm s}(z_{\rm m})$ and $D_{\rm v}(z_{\rm m})\omega_{\rm m}$. 
The interplay between the constraints on these parameter combinations shapes the allowed region in the 
$\omega_{\rm m}-D_{\rm V}$ plane. The dotted line in Fig.~\ref{fig:om_dv} corresponds to 
a constant value of the quantity \citep{Eisenstein2005}
\begin{equation}
 A(z_{\rm m})=D_{\rm v}(z_{\rm m})\frac{\sqrt{\Omega_{\rm m}H^2_0}}{cz_{\rm m}},
\label{eq:apar}
\end{equation}
which approximately describes the resulting degeneracy between $\omega_{\rm m}$ and $D_{\rm V}(z_{\rm m})$. 
To take into account the effect of the baryon density on the scale $k_{\rm eq}$, this quantity
should be defined in terms of the shape parameter $\Gamma$. However, we maintain the usual
definition to simplify the comparison with previous analyses. The CMASS correlation function implies a 
constraint of $A(z_{\rm m})=0.444\pm0.014$.

Two of our companion papers, \citet{Blanton2012} and \citet{Reid2012}, 
study the cosmological implications of the galaxy clustering in the CMASS sample.
While \citet{Blanton2012} is based on the constraints inferred
from the BAO signal, \citet{Reid2012} deals with the analysis of redshift-space distortions.
Both of these studies present constraints on the quantity
\begin{equation}
 \alpha=y_{\rm s}^{\rm fid}(z_{\rm m})/y_{\rm s}(z_{\rm m}),
\end{equation}
where $y_{\rm s}^{\rm fid}(z_{\rm m})$ is the value corresponding to our fiducial cosmology.
Dropping the priors on $\omega_{\rm b}$ and $n_{\rm s}$, we obtain the constraint 
$y_{\rm s}(z_{\rm m})=0.0745\pm0.0014$, which implies $\alpha=1.015 \pm 0.019$. This result  
is in good agreement with the constraints reported in our companion papers:
\citet{Reid2012} obtain $\alpha=1.023\pm0.019$, while \citet{Blanton2012} find $\alpha=1.016\pm0.017$
from the pre-reconstruction correlation function, and a post-reconstruction``consensus'' value 
between $\xi(s)$ and $P(k)$ of $\alpha=1.033\pm0.017$. This agreement is a clear demonstration of
the consistency between the different analysis techniques implemented in these studies.

\section{Cosmological implications}
\label{sec:results}

In this Section, we perform a systematic study of the constraints placed on the values 
of the cosmological parameters described Section~\ref{ssec:param}.
In Section~\ref{ssec:lcdm}, we present the results for 
the simple $\Lambda$CDM cosmological model with six free parameters.
In Section~\ref{ssec:omk} we discuss our constraints on non-flat models.
Section~\ref{ssec:fnu} deals with the constraints on 
the fraction of massive neutrinos.
In Section~\ref{ssec:tensor} we allow for non-zero tensor modes. 
In Section~\ref{ssec:darkenergy} we focus on the constraints on the nature of dark energy.  
Models where the dark energy equation of state is constant over time are analysed in Section~\ref{ssec:wde},
while Section~\ref{ssec:wa} explores the constraints on the redshift dependence of $w_{\rm DE}$, 
parametrized according to equation~(\ref{eq:wa}). Finally, Section~\ref{ssec:wok} 
shows the impact of allowing also for 
models with $\Omega_k\neq0$ on the constraints on $w_{\rm DE}$. 
Tables~\ref{tab:lcdm}-\ref{tab:wok} in Appendix~\ref{sec:tables}
summarize the constraints obtained in these parameter spaces using different combinations of 
the datasets described in Sections~\ref{sec:corfunc} and \ref{sec:moredata}.

\subsection{The $\Lambda$CDM model}
\label{ssec:lcdm}

In this Section we focus on the $\Lambda$CDM model and discuss the constraints on the parameter space of 
equation~(\ref{eq:param}). 
The CMB data alone are able to provide tight constraints on this parameter space, especially 
on quantities such as $\omega_{\rm b}$, $\theta$ and $\tau$, whose constraints show almost no
variation when other datasets are included in the analysis. 
However, the constraints on other parameters are improved by considering additional datasets.

\begin{figure}
\centering
\centerline{\includegraphics[width=\columnwidth]{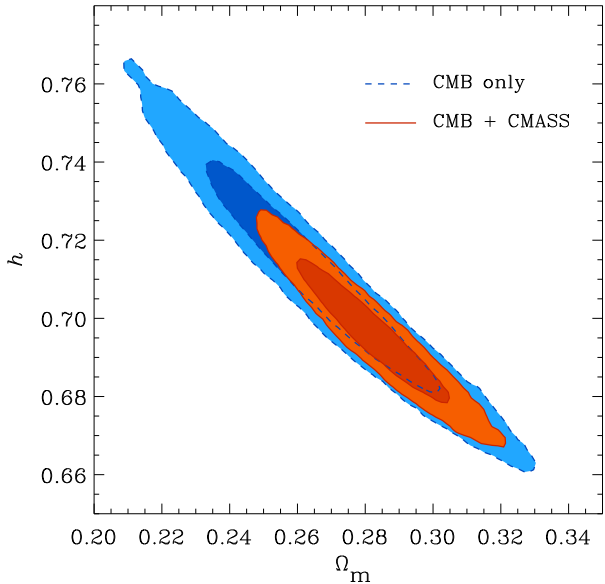}}
\caption{
The marginalized constraints in the $\Omega_{\rm m}-h$ plane for the $\Lambda$CDM 
parameter set. The dashed lines show the 68 and 95 per cent contours obtained using CMB information alone. The solid 
contours correspond to the results obtained from the combination of CMB data plus the shape 
of the CMASS $\xi(s)$. 
}
\label{fig:om_h}
\end{figure}

The dashed lines in Fig.~\ref{fig:om_h} show the two-dimensional marginalized constraints in the 
$\Omega_{\rm m}-h$ plane obtained using CMB data alone. The contours show a degeneracy that follows
approximately a line of constant $\Omega_{\rm m}h^3$ \citep{Percival2002}.
This degeneracy limits the accuracy of the one-dimensional constraints on these parameters, which from the CMB 
data alone are $\Omega_{\rm m}=0.266\pm0.024$ and $h=0.710\pm 0.020$.
The solid lines in Fig.~\ref{fig:om_h} show the result of combining the CMB 
measurements with the CMASS correlation function. The extra information contained in the shape of $\xi(s)$
partially breaks this degeneracy, leading to tighter constraints of $\Omega_{\rm m}=0.282 \pm 0.015$
and $h=0.696\pm0.012$. The dashed line in Fig.~\ref{fig:corfunc} corresponds to the best
fitting model obtained in this case. This model gives an excellent match to both the location of the BAO peak
and the full shape of the CMASS correlation function. On scales $s > 80\,h^{-1}{\rm Mpc}$,
the model slightly under-predicts the amplitude of $\xi(s)$. Note, however, that on these scales the individual
points in the measurement are correlated.
Taking into account the full covariance matrix, this model gives $\chi^2=27$ for
32 degrees of freedom, providing an excellent fit. 
This model requires a real-space bias factor
\citep[i.e., computed after accounting for the boost factor of][]{Kaiser1987} of
$b_{\rm r}=1.96\pm0.09$. This value is in excellent agreement with the results of \citep{Nuza2012},
who estimated a bias factor of $b_{\rm r}\simeq2$ from an abundance matching analysis of the small and
intermediate scale clustering of the CMASS sample based on the Multidark simulation.

\begin{figure}
\centering
\centerline{\includegraphics[width=\columnwidth]{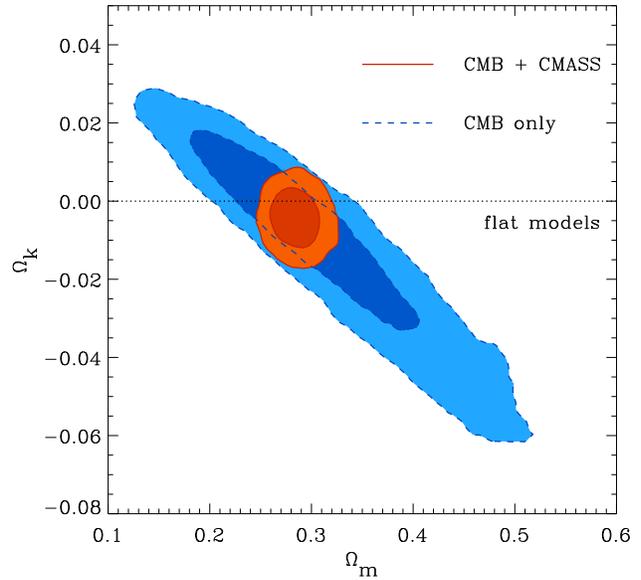}}
\caption{
The marginalized posterior distribution in the $\Omega_{\rm m}-\Omega_{k}$ plane for the $\Lambda$CDM 
parameter set extended to allow for non-flat models.
The dashed lines show the 68 and 95 per cent contours obtained using CMB information alone. The solid 
contours correspond to the results obtained from the combination of CMB data plus the shape 
of the CMASS $\xi(s)$. The dotted line corresponds to the $\Lambda$CDM model, where $\Omega_{k}=0$.
}
\label{fig:om_ok}
\end{figure}

The results presented here are completely consistent with those of 
\citet{Blanton2012}, who explored the cosmological implications of the BAO signal in the CMASS
correlation function. From the combination of this information with the latest data from the
WMAP satellite, they find $\Omega_{\rm m}= 0.298\pm0.017$ and $h=0.684\pm0.013$ when the parameter
space is restricted to the $\Lambda$CDM model. Although this agreement is not surprising, as the
two analyses are based on the same galaxy sample,
it is a clear indication of the consistency between the two analysis techniques.

\begin{figure*}
\centering
\centerline{\includegraphics[width=0.9\textwidth]{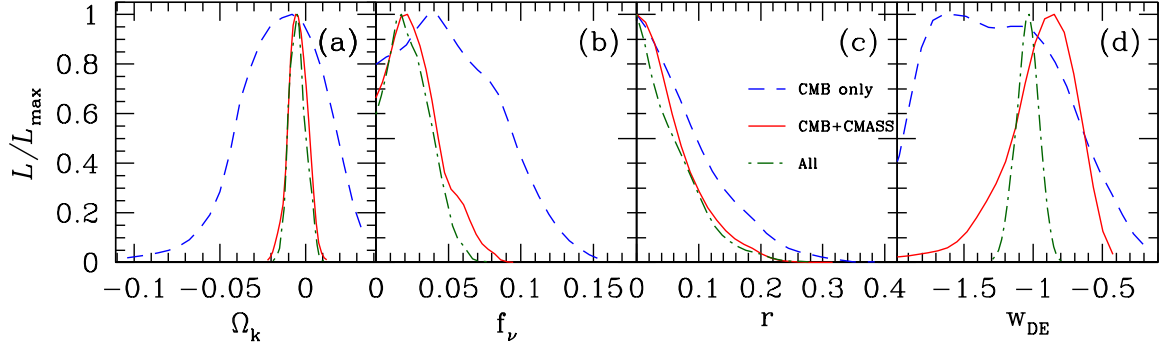}}
\caption{
The marginalized, one-dimensional likelihood distribution of the extensions of the $\Lambda$CDM model explored in Sections~\ref{ssec:omk} through \ref{ssec:darkenergy}.
Dashed lines indicate the constraints obtained from CMB information only, solid lines correspond to the results of
the CMB plus the shape of the CMASS $\xi(s)$, and the
dot-dashed lines show full constraints including also BAO and SN data. 
}
\label{fig:extensions}
\end{figure*}

  Although consistent within 1~$\sigma$, the CMASS correlation function prefers somewhat higher
values of $\Omega_{\rm m}$ than the CMB data. This difference can be traced back to the values of
$y_{\rm s}(z_{\rm m})$ obtained from these datasets individually. In the $\Lambda$CDM parameter space it is
possible to obtain a constraint on this quantity on the basis of CMB information alone. In this case we obtain
$y_{\rm s}(z_{\rm m})=0.0762\pm0.0018$, while the CMASS $\xi(s)$ gives $y_{\rm s}(z_{\rm m})=0.0742\pm0.0014$.
We will return to this point in Section~\ref{ssec:northsouth}, where we analyse the clustering properties of the
NGC and SGC sub-samples separately. 

  As can be seen in Figure~\ref{fig:om_h}, by preferring higher values of $\Omega_{\rm m}$, the CMASS 
correlation function also leads to slightly lower values of the Hubble parameter than in the CMB only case.
Although this value is lower than the direct measurement of \citet{Riess2011}, the difference is not statistically
significant. As discussed in \citet{Blanton2012} and \citet{Mehta2012} this difference can be reduced if the
effective number of relativistic species, $N_{\rm eff}$, is allowed to deviate from the standard value of
$N_{\rm eff}=3.04$.

As shown in Table~\ref{tab:lcdm}, when the SN and BAO data are added to the analysis, 
the results point towards values of $\Omega_{\rm m}$ similar to those of the CMB+CMASS case.
Combining the information from all these datasets,
the recovered values of $\Omega_{\rm m}$ and $h$ are similar to the CMB+CMASS results and the
uncertainties are reduced by 33\%. In this case we find 
$\Omega_{\rm m}= 0.2846_{-0.0097}^{+0.0095}$ and $h=0.6941\pm0.0081$.

Recent analyses have consistently shown evidence of a departure from the scale-invariant 
primordial power spectrum of scalar fluctuations \citep{Sanchez2006,Spergel2007,Komatsu2009,Komatsu2011,Keisler2011}.
Our CMB+CMASS constraint on the spectral index is $n_{\rm s}=0.9620_{-0.0091}^{+0.0093}$, increasing the 
significance of this detection to 4.1~$\sigma$. This limit is almost unchanged when all datasets are considered,
in which case we get $n_{\rm s}=0.9613_{-0.0090}^{+0.0089}$.
The deviation from scale-invariance of the primordial power spectrum has important implications, as most
inflationary models predict that the scalar spectral index is less than one \citep{Linde2008}.
However, these models also predict the presence of non-zero tensor primordial fluctuations.
As we will see in Section~\ref{ssec:tensor}, although the constraints on $n_{\rm s}$ become weaker
when the tensor-to-scalar ratio, $r$, is allowed to vary, we also detect a deviation from scale invariance
at the 99.7 per cent confidence level (CL) in this case.

The results from our study show that the standard $\Lambda$CDM model is able to accurately describe all the datasets
that we have included in our analysis and that the values of its basic parameters are constrained
to an accuracy higher than 5 per cent. 
In the following sections we focus on constraining possible deviations from this simple model.

\subsection{Non-flat models}
\label{ssec:omk}

In this Section we drop the assumption of a flat Universe and allow for
models where $\Omega_k\neq0$.
This parameter space is poorly constrained by the CMB data due to the 
so-called geometrical degeneracy \citep{Efstathiou1999} relating the physical size of the sound horizon
at recombination $r_{\rm s}(z_*)$, and the angular diameter distance $D_{\rm A}(z_*)$.
The former determines the true physical scale of the acoustic oscillations, while 
the later controls its mapping onto angular scales in the sky.
Models with the same value of $\Theta=r_{\rm s}(z_*)/D_{\rm A}(z_*)$ predict the same position 
of the acoustic peaks in the CMB spectrum and cannot be distinguished on the basis of the primary CMB fluctuations alone. 
This degeneracy is shown by the dashed lines in Fig.~\ref{fig:om_ok}, which correspond to
the 68 and 95 per cent CL contours in the $\Omega_{\rm m}-\Omega_{k}$ plane obtained from the CMB data.
The dashed line in panel (a) of Fig.~\ref{fig:extensions} shows the corresponding marginalized constraints on $\Omega_{k}$,
which allow for significant deviations from the $\Lambda$CDM model value.
In this case we obtain $\Omega_{k}=-0.014_{-0.025}^{+0.022}$ and $\Omega_{\rm m}=0.32_{-0.09}^{+0.10}$. 

As shown by the solid lines in Fig.~\ref{fig:om_ok}, the constraints on $y_{\rm s}(z_{\rm m})$ and $A(z_{\rm m})$
provided by the CMASS correlation function are very effective at breaking this degeneracy, leading to a drastic
decrease in the range of allowed values for these parameters. 
The solid line in panel (a) of Fig.~\ref{fig:extensions} corresponds to the posterior distribution of $\Omega_{k}$ obtained from the CMB+CMASS combination, 
which is in much closer agreement with a flat universe. 
In this case we obtain $\Omega_{\rm m}=0.285\pm0.015$ and $\Omega_{k}=-0.0043\pm0.0049$.

\citet{Blanton2012} explored the same parameter space using the CMASS BAO signal.
From the combination of this measurement with WMAP data they find $\Omega_{\rm m}= 0.299\pm0.016$ and $\Omega_{k}=-0.008\pm0.005$.
These constraints are in good agreement with findings reported here, although they show a preference for slightly higher values
of the matter density parameter.

The inclusion of the SN and BAO datasets does not significantly improve the results over those obtained using the CMB+CMASS combination,
with a final constraint of $\Omega_{k}=-0.0045\pm0.0042$ obtained from the combination of all datasets. 
This means that current observations restrict possible variations in the spatial curvature of the 
Universe up to a level of $\Delta \Omega_k\simeq 4 \times 10^{-3}$. 

\subsection{Massive neutrinos}
\label{ssec:fnu}

In the standard $\Lambda$CDM scenario the dark matter component is given entirely by cold dark matter. 
However, over the last decade a number of experiments have shown clear evidence of neutrino oscillations,
implying that the three known types of neutrino have a non-zero mass and contribute to the 
total energy budget of the Universe. 
These observations are only sensitive to the mass-squared differences between neutrino
flavours rather than on their absolute masses.
Absolute neutrino mass measurements can be obtained from tritium $\beta$-decay experiments,
which at present provide upper limits of
$\sum m_{\nu} < 6\,{\rm eV}$ at the 95 per cent CL \citep{Lobashev2003,Eitel2005,Lesgourgues2006}.
Future experiments like KATRIN are expected to improve these bounds by an order of magnitude
 \citep{Otten2008}.
Until then, the best observational window into neutrino masses is provided by cosmological observations,
in particular by the combination of CMB and LSS datasets
\citep{Hu1998, Elgaroy2002, Hannestad2002, Sanchez2006, Reid2010b, Deputter2012}.
A variation in the neutrino mass can alter the redshift of matter-radiation equality, thereby affecting the 
CMB power spectrum. Additionally, until the time when they become non-relativistic, neutrinos free-stream out
of density perturbations, suppressing the growth of structures on scales smaller than the horizon at that
time, which is a function of their mass. This affects the shape of the matter power spectrum and the correlation function.

In this Section we explore the constraints on the neutrino fraction, $f_{\nu}$.
As current estimates of the differences in the neutrino mass hierarchy are an order of magnitude lower than 
the constraints on $\sum m_{\nu}$ from cosmological observations, these are not yet sensitive to the masses of
individual neutrino eigenvalues; we therefore assume three neutrino species of equal mass.
The dashed line in panel (b) of Fig.~\ref{fig:extensions} corresponds to the constraints on the neutrino
fraction obtained from CMB data alone. In this case, we find $f_{\nu} < 0.11$ at 95 per cent CL. 
The solid line in the same panel shows the effect of including 
also the information from the shape of the CMASS correlation function, which drastically reduces this limit to 
$f_{\nu} < 0.055$ at 95 per cent CL.
Our results can be converted into constraints on the sum of the three neutrino masses using
equation~(\ref{eq:mnu}) to obtain
$\sum m_{\nu}< 1.4\,{\rm eV}$ (95 per cent CL) in the CMB only case, and
$\sum m_{\nu}< 0.61\,{\rm eV}$ (95 per cent CL) for CMB data plus
the CMASS $\xi(s)$.

\begin{figure}
\centering
\centerline{\includegraphics[width=\columnwidth]{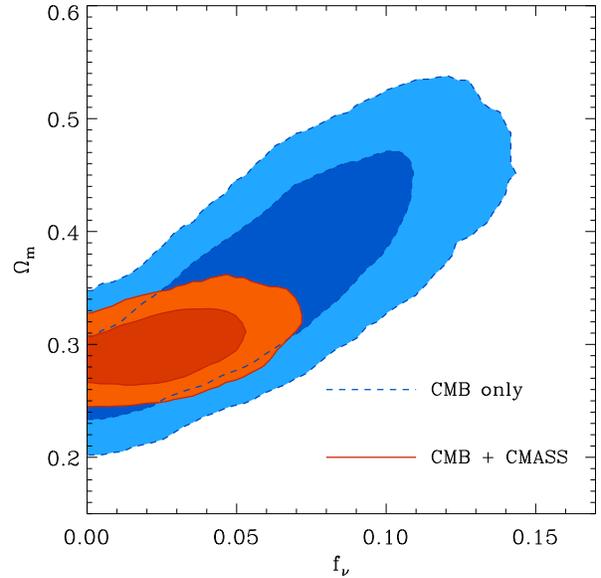}}
\caption{
The marginalized posterior distribution in the $f_{\nu}-\Omega_{\rm m}$ plane for the $\Lambda$CDM 
parameter set extended by allowing for a non-negligible fraction of massive neutrinos.
The dashed lines show the 68 and 95 per cent contours obtained using CMB information alone. The solid 
contours correspond to the results obtained from the combination of CMB data plus the shape 
of the CMASS $\xi(s)$.
}
\label{fig:fnu_om}
\end{figure}

Fig.~\ref{fig:fnu_om} shows the 68\% and 95\% constraints in the $\Omega_{\rm m}-f_{\nu}$ plane. 
As shown by the dashed lines, when CMB data alone is considered, allowing for $f_{\nu}\neq 0$ leads to 
significantly weaker constraints on $\Omega_{\rm m}$ with respect to the $\Lambda$CDM case, with its 
range of allowed values increasing by more than a factor of two. 
The information in the shape of the CMASS correlation function improves these constraints, 
leading to $\Omega_{\rm m}= 0.298\pm0.019$, with a similar accuracy to that of the $f_{\nu}=0$ case.

In a recent analysis, \citet{Deputter2012} explored the constraints on $\sum m_{\nu}$ 
from the angular power spectrum of a galaxy sample drawn from BOSS-DR8 following the CMASS 
selection criteria, as measured by \citet{Ho2012}.
From the combination of this measurement with WMAP7 information, \citet{Deputter2012}
obtained a limit of $\sum m_{\nu}< 0.56\,{\rm eV}$ at 95\% CL, which is relaxed to
$\sum m_{\nu}< 0.90\,{\rm eV}$ (95\% CL) when a more conservative galaxy bias model is implemented.
The similarity between these limits and our CMB+CMASS constraint illustrates the power of using 
the full three dimensional clustering information, which can compensate for the much larger volume 
probed by the sample analysed by \citet{Deputter2012}.

Although not directly sensitive to $f_{\nu}$, the additional information from SN or BAO measurements
improves the limits on the neutrino fraction by constraining parameters which are 
degenerate with this quantity. Combining all datasets we obtain $f_{\nu}< 0.049$ and 
$\sum m_{\nu}< 0.51\,{\rm eV}$ at 95 per cent CL. 
In the analysis of \citet{Deputter2012}, the inclusion of the SN and $H_0$ measurements provided a
tighter constraint, with $\sum m_{\nu}< 0.26\,{\rm eV}$ at 95\% CL and $\sum m_{\nu}< 0.36\,{\rm eV}$ (95\% CL) 
for the two galaxy bias models they analysed.

An extension of the current analysis to include information from $\xi(s)$ on smaller scales, where 
it is more sensitive to the effect of neutrino free-streaming, could help to improve the constraints on the 
neutrino fraction even further. However, as pointed out by \citet*{Swanson2010}, effects related to 
non-linearities and galaxy bias on these scales might impose a limitation on the robustness of clustering
measurements as a means to obtain bounds on the neutrino mass.
For this reason, the constraints on $\sum m_{\nu}$ presented here should be regarded as conservative,
while the full constraining power of the CMASS sample on this quantity will be explored in future studies.

\subsection{Tensor modes}
\label{ssec:tensor}

We now extend the parameter space of equation~(\ref{eq:param}) to include the tensor-to-scalar amplitude
ratio $r$. This is the parameter space most relevant for the study of inflation as the most simple
inflationary models predict non-zero primordial tensor modes
\citep[i.e. gravitational waves,][]{ Linde2008}.

\begin{figure}
\centering
\centerline{\includegraphics[width=\columnwidth]{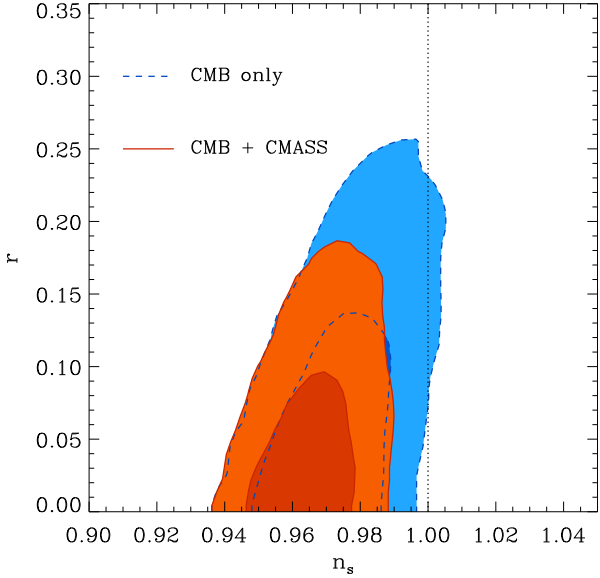}}
\caption{
The marginalized posterior distribution in the $n_{\rm s}-r$ plane for the $\Lambda$CDM 
parameter set extended by allowing for non-zero primordial tensor modes.
The dashed lines show the 68 and 95 per cent contours obtained using CMB information alone. The solid 
contours correspond to the results obtained from the combination of CMB data plus the shape 
of the CMASS $\xi(s)$. The dotted line corresponds to the scale-invariant scalar primordial 
power spectrum, with $n_{\rm s}=1$.
}
\label{fig:ns_r}
\end{figure}

Panel (c) of Fig.~\ref{fig:extensions} shows the marginalized constraints on $r$
for the cases of CMB data only (dashed lines) and CMB plus the CMASS $\xi(s)$ (solid lines).
The constraints imposed on $r$ by CMB information alone are $r < 0.21$ (95 per cent CL).
The CMASS correlation function tightens this limit to $r<0.16$ at the 95 per cent CL.
This result is only marginally improved by 
the additional information of the SN and BAO datasets to 
our final constraint of $r<0.15$ (95 per cent CL). These results 
show good agreement with the constraints of \citet{Keisler2011}, who found $r<0.17$ (95 per cent CL)
from the combination of the same CMB datasets with BAO and $H_0$ measurements. 

Fig.~\ref{fig:ns_r} shows the likelihood contours in the $n_{\rm s}-r$ plane obtained by means of
CMB data alone (dashed lines),
and its combination with the CMASS $\xi(s)$ (solid lines). 
Tensor modes contribute to the CMB temperature power spectrum only on large angular scales ($\ell < 400$). 
An increase in the value of $r$ can be compensated for by reducing the amplitude of the scalar modes,
thereby maintaining the total amplitude of the temperature fluctuations at a constant level. The consequent
decrease of power on smaller angular scales can be compensated for by increasing in the scalar spectral index,
$n_{\rm s}$. Although, as discussed in \citet{Keisler2011}, the information from the small angular scales of
the CMB fluctuations provided by SPT does a good job at breaking this degeneracy, a residual relation between
these parameters limits the accuracy of their marginalized constraints. By also including the information
from the shape of the CMASS
correlation function, it is possible to restrict the range of allowed values for these parameters even further.
In particular, this combination allows us to detect a deviation from the scale-invariant primordial power
spectrum (indicated by the vertical dotted line) with $n_{\rm s}<1$ at the 99.7\% CL, even in the presence of
tensor modes. This detection has strong implications for the inflationary paradigm.

\begin{figure}
\centering
\centerline{\includegraphics[width=\columnwidth]{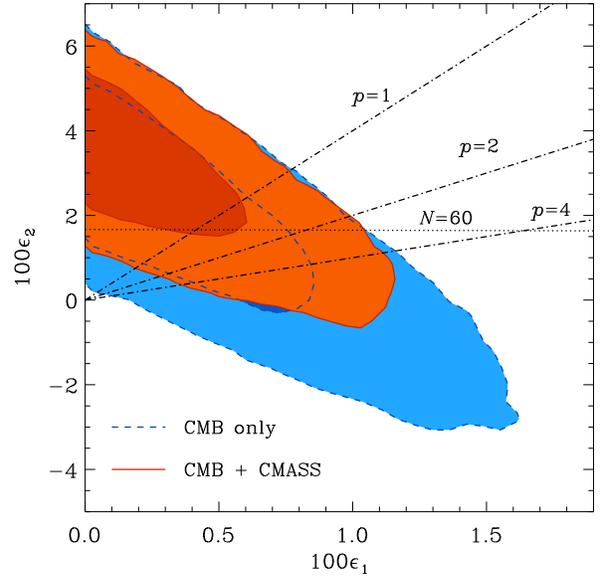}}
\caption{ 
The marginalized posterior distribution in the $\epsilon_1-\epsilon_2$ plane for the $\Lambda$CDM 
parameter set extended by allowing for non-zero primordial tensor modes.
The dashed lines show the 68 and 95 per cent contours obtained using CMB information alone. The solid 
contours correspond to the results obtained from the combination of CMB data plus the shape 
of the CMASS $\xi(s)$. 
The dot-dashed lines correspond to chaotic inflationary models with $p=1$, 2 and 4,
as indicated by the labels. The dotted line corresponds to a constant value of $N=60$.
}
\label{fig:e1_e2}
\end{figure}

We can explore the implications of our results in terms of constraints on inflationary models
by analysing the horizon flow parameters of \citet{Schwarz2001}. These are a hierarchy of parameters describing the 
evolution of the Hubble factor during inflation. The first parameter is given by
$\epsilon_1\equiv -{\rm d}\ln H(N)/{\rm d}N$, where $N$ is the 
number of e-foldings before the end of inflation at which our pivot scale crosses the Hubble radius during inflation, and the remaining ones are defined through the relation
\begin{equation}
\epsilon_{j+1}\equiv\frac{{\rm d}\ln |\epsilon_{j}|}{{\rm d}N},\, j\ge 1.
 \label{eq:ei}
\end{equation}
The weak energy condition implies that $\epsilon_1>0$, while a necessary condition for inflation is
$\epsilon_1<1$ (which implies $\ddot{a}>0$).
The slow-roll approximation can be expressed as $|\epsilon_{j}|\ll1$, for all $j>0$. In this limit,
these parameters satisfy the relations
\begin{eqnarray}
r &=& 16\epsilon_1, \\
n_{\rm s} &=& 1-2\epsilon_1-\epsilon_2.
\label{eq:inflation1}
\end{eqnarray}
These relations can be used to translate our constraints on $n_{\rm s}$ and $r$ into the
$\epsilon_1-\epsilon_2$ plane. Fig.~\ref{fig:e1_e2} shows the constraints obtained in this way.  
Marginalizing over $\epsilon_2$, the combination of CMB data plus the CMASS correlation function
implies the limit $\epsilon_1<0.0097$ at the 95 per cent CL. 
These datasets strongly favour models with positive values of $\epsilon_2$, in which inflation
will end naturally with a violation of the slow-roll approximation \citep{Leach2003}. 
From the CMB+CMASS combination we obtain the limit $\epsilon_2>0$ at the 95.8 per cent CL, which 
is only marginally improved to the 97 per cent CL with the inclusion of the SN and additional
BAO measurements.

The horizon flow parameters are related to the inflaton potential $V$ and its derivatives
with respect to the inflaton field $\phi$. Then, they can be used to constrain which type of
potentials are compatible with the observations \citep[see e.g.][]{Liddle2003a,Kinney2008,Finelli2010}.
As an example, we explore the constraints on a particular
class of models, that of the chaotic (or monomial) inflation, in which the inflationary
phase is driven by a potential of the form $V(\phi)\propto \phi^p$.
These models predict a simple relation between the horizon flow parameters, the power-law
index, $p$, and the number of e-folds, $N$, given by \citep{Leach2003}:
\begin{eqnarray}
\epsilon_2 &=& \frac{4}{p}\epsilon_1, \label{eq:inflation2}\\
N &=& \frac{p}{4}\left(\frac{1}{\epsilon_1}-1\right).
\label{eq:inflation3}
\end{eqnarray}
The dot-dashed lines in Fig.~\ref{fig:e1_e2} correspond to chaotic inflationary models with $p=1$, 2 and 4,
as indicated by the labels.
As can be seen from equations~(\ref{eq:inflation2}) and (\ref{eq:inflation3}), a given value of $N$
corresponds approximately to a constant value of $\epsilon_2$. 
For the pivot scale considered here, a plausible upper limit for the number of e-folds is
$N\lesssim60$ \citep{Dodelson2003,Liddle2003b}, corresponding to $\epsilon_2\gtrsim0.017$ 
(indicated by the dotted line in Fig.~\ref{fig:e1_e2}).
If we restrict our analysis to this region of the parameter space, we see that 
models with $p\gtrsim2$ are strongly disfavoured by the data.
In fact, the marginalized distribution for $p$ obtained from the CMB+CMASS combination after applying
this prior implies a limit of $p<1.2$ at the 95 per cent CL, imposing a strong constraint on the viable chaotic
inflationary models.

\subsection{The dark energy equation of state}
\label{ssec:darkenergy}

Until now we have assumed that the dark energy component corresponds
to a cosmological constant, with a fixed equation of state specified by $w_{\rm DE} = -1$. 
In this Section, we allow for more general dark energy models. In Section ~\ref{ssec:wde} we explore the constraints 
on the value of $w_{\rm DE}$ (assumed redshift-independent). In Section ~\ref{ssec:wa} we 
obtain constraints on the time evolution of this parameter, parametrized according to equation~(\ref{eq:wa}). 
Section~\ref{ssec:wok} deals with the effect of the assumption of a flat universe on the 
constraints on $w_{\rm DE}$.

 In these tests we consider models with $w_{\rm DE} < -1$, corresponding to phantom energy
\citep[see][and references therein]{Copeland2006}. When exploring constraints on dynamical dark
energy models, these are allowed to cross the so-called phantom divide, $w_{\rm DE} = -1$.  
In the framework of general relativity, a single fluid, or a single scalar field without
higher derivatives, cannot cross this threshold since it would become
gravitationally unstable \citep{Feng2004,Vikman2005,Hu2005,Xia2008},
requiring at least one extra degree of freedom.  However, models with
more degrees of freedom are difficult to implement in general dark
energy studies.  Here we follow the parametrized post-Friedmann (PPF)
approach of \citet{Fang2008}, as implemented in {\sc CAMB}, which
provides a simple solution to these problems for models in which the
dark energy component is smooth compared to the dark matter.
Alternatively, as proposed by \citet{Zhao2005}, it is possible to
consider the dark energy perturbations using a two-field model, with
one of the fields being quintessence-like and the other one phantom-like
\citep[e.g. the quintom model proposed in][]{Feng2004} without
introducing new internal degrees of freedom. 
Both approaches give consistent results.

\subsubsection{Time-independent dark energy equation of state}
\label{ssec:wde}

\begin{figure}
\centering
\centerline{\includegraphics[width=\columnwidth]{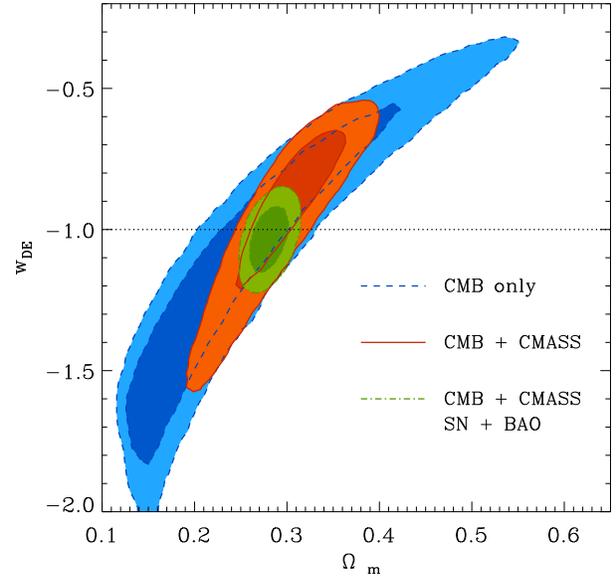}}
\caption{
The marginalized posterior distribution in the $\Omega_{\rm m}-w_{\rm DE}$ plane for the $\Lambda$CDM 
parameter set extended by including the redshift-independent value of $w_{\rm DE}$ as an additional parameter.
The dashed lines show the 68 and 95 per cent contours obtained using CMB information alone. The solid 
contours correspond to the results obtained from the combination of CMB data plus the shape of the CMASS $\xi(s)$.
The dot-dashed lines indicate the results obtained from the full dataset combination (CMB+CMASS+SN+BAO).
The dotted line corresponds to the $\Lambda$CDM model, where $w_{\rm DE}=-1$.
}
\label{fig:wde}
\end{figure}

In this Section we explore the constraints on the parameter set of equation~(\ref{eq:param})
extended by including the redshift-independent value of $w_{\rm DE}$ as an additional parameter.
The dashed lines in Fig.~\ref{fig:wde} show the two-dimensional marginalized constraints in 
the $\Omega_{\rm m}-w_{\rm DE}$ plane obtained from CMB data alone. There is a strong degeneracy between these
parameters along which different models predict the same angular position for the peaks in the CMB power 
spectrum. This is analogous to the geometrical degeneracy described in Section~\ref{ssec:omk},
corresponding to models with constant values of $\Theta$.
This degeneracy leads to poor one-dimensional constraints of $w_{\rm DE}=-1.15_{-0.39}^{+0.39}$ and 
$\Omega_{\rm m}=0.248_{-0.088}^{+0.093}$. 

The solid lines in Fig.~\ref{fig:wde} show the effect of including the CMASS correlation function
in the analysis. The constant-$\Theta$ degeneracy can be partially broken by providing an additional distance
constraint. The constraint on $y_{\rm s}(z_{\rm m})$ provided by $\xi(s)$ breaks the degeneracy between
$\Omega_{\rm m}$ and $w_{\rm DE}$, tightening the constraints on the dark energy equation of state.
The impact  of including the CMASS correlation function on the marginalized constraints on $w_{\rm DE}$
can be seen in panel (d) of Fig.~\ref{fig:extensions} where the dashed lines correspond to the result of the
CMB only case and the solid lines the one of the CMB+CMASS combination.  
In this case we obtain $\Omega_{\rm m}=0.295_{-0.042}^{+0.041}$ and $w_{\rm DE}=-0.95_{-0.20}^{+0.21}$, 
in good agreement with a cosmological constant. 

From the combination of the BAO signal inferred from the CMASS $P(k)$ and $\xi(s)$ with WMAP data, 
\citet{Blanton2012} obtained the constraints $\Omega_{\rm m}=0.323\pm0.043$ and $w_{\rm DE}=-0.87\pm0.24$,
in good agreement with our findings. As in the previous parameter spaces, this is a clear indication of the 
consistency between the two analysis techniques. The extra information in the shape of 
$\xi(s)$ improves the constraints on the dark energy equation of state by $\sim$20\% with respect to
the BAO only result, indicating that, at this redshift, most of the information on this parameter
is obtained through the measurement of $y_{\rm s}$.

In a recent analysis, \citet{Montesano2012} used the full shape of the power spectrum of a sample of LRGs from the
final SDSS-II, combined with a compilation of CMB experiments, to obtain the constraint $w_{\rm DE}=-1.02\pm0.13$. 
\citet{Mehta2012} combined the BAO distance measurement derived by \cite{Padmanabhan2012} and 
\cite{Xu2012} from the same galaxy sample with WMAP data, to obtain $w_{\rm DE}=-0.92\pm0.13$. 
As these measurements are based on observations are lower redshifts, which are more sensitive to variations 
in $w_{\rm DE}$, they provide slightly tighter constraints on this parameter than the CMB+CMASS combination.

Including also the additional BAO data in the analysis gives similar results to the CMB+CMASS case, with a
constraint on the dark energy equation of state of $w_{\rm DE}=-0.91_{-0.11}^{+0.11}$. 
When the SN data is considered in the analysis instead of the BAO, the resulting constraints are in better agreement with the standard $\Lambda$CDM value, with $w_{\rm DE}=-1.054_{-0.076}^{+0.077}$.
It is interesting to note that this result is mostly driven by the CMASS+SN combination. In fact, 
the combined information from these two datasets provides the constraint $w_{\rm DE}=-1.04\pm0.11$,
independently of any CMB data.
Our final constraints obtained from the combination of all datasets are shown by the dot-dashed lines in 
Fig.~\ref{fig:wde}, corresponding to  $\Omega_{\rm m}=0.281\pm0.012$ and $w_{\rm DE}=-1.033_{-0.073}^{+0.074}$.
This result is in excellent agreement with the standard $\Lambda$CDM model value of $w_{\rm DE}=-1$,
indicated by a dotted line in Fig.\ref{fig:wde}.

\subsubsection{The time evolution of $w_{\rm DE}$}
\label{ssec:wa}

\begin{figure}
\centering
\centerline{\includegraphics[width=\columnwidth]{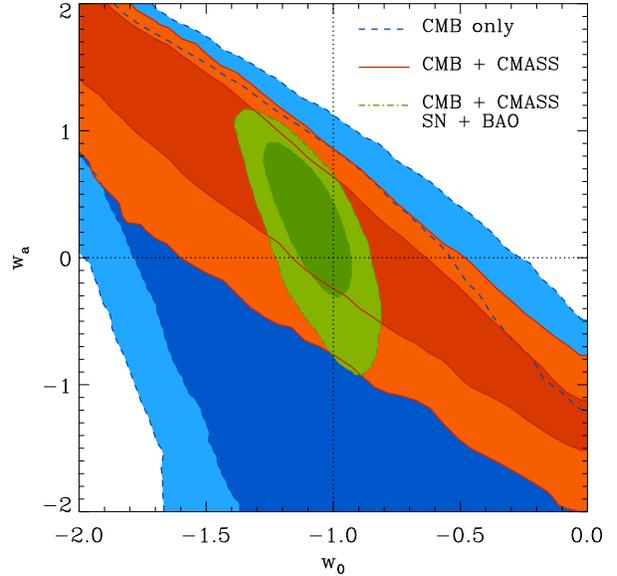}}
\caption{
The marginalized posterior distribution in the $w_0-w_a$ plane for the $\Lambda$CDM 
parameter set extended by allowing for variations on $w_{\rm DE}(a)$, parametrized as in equation~(\ref{eq:wa}).
The dashed lines show the 68 and 95 per cent contours obtained using CMB information alone. The solid 
contours correspond to the results obtained from the combination of CMB data plus the shape of the CMASS $\xi(s)$.
The dot-dashed lines indicate the results obtained from the full dataset combination (CMB+CMASS+SN+BAO).
The dotted lines correspond to the canonical values in the $\Lambda$CDM model.
}
\label{fig:bwa_w0_wa}
\end{figure}

In the $\Lambda$CDM model, the equation of state parameter is characterized by the fixed value $w_{\rm DE}=-1$
at all times.
A detection of a deviation from this prediction would be a clear signature of the need of alternative 
dark energy models. In this Section, we explore the constraints on the redshift dependence of $w_{\rm DE}$
which we parametrize according to equation~(\ref{eq:wa}). 

The dashed lines in Fig.~\ref{fig:bwa_w0_wa} show the two-dimensional marginalized constraints in the $w_0-w_a$
plane obtained from the CMB data alone. 
This case provides only weak constraints on these parameters, allowing for models
where the value of $w_{\rm DE}$ can vary significantly over time. The inclusion of the CMASS correlation function reduces this allowed region 
to a linear degeneracy between $w_0$ and $w_a$ which can still accommodate large deviations from the $\Lambda$CDM values,
indicated by the dotted lines. At least a third dataset is required to obtain more restrictive constraints. 
In the CMB+CMASS+SN case, we obtain $w_0=-1.09\pm0.11$ and $w_a=0.12_{-0.47}^{+0.48}$, that change to 
$w_0=-0.95\pm0.27$ and $w_a=0.05_{-0.61}^{+0.62}$ if the SN data are replaced by the additional BAO measurements.
The dot-dashed lines in Fig.~\ref{fig:bwa_w0_wa} correspond to our tightest constraints, obtained by combining all datasets,
where we obtain the marginalized values $w_0=-1.08\pm0.11$ and $w_a=0.23\pm0.42$.

A useful quantity to characterize the constraints on the redshift evolution of the dark energy equation of state 
is the pivot redshift, $z_{\rm p}$, defined as the point where the uncertainty on $w_{\rm DE}(a)$ is minimized
\citep{Huterer2001,Hu2004,Albrecht2006}.
The parametrization of equation~(\ref{eq:wa}) implies that this redshift corresponds to the scale factor
\begin{equation}
a_{\rm p}=1+\frac{\langle \delta w_{\rm 0}\delta w_{\rm a} \rangle}{\langle\delta w_{\rm a}^2 \rangle}.
\label{eq:pivot}
\end{equation}
The corresponding pivot redshift for the CMB+CMASS combination is given by $z_{\rm p}=1.21$, for which we obtain
$w_{\rm DE}(z_{\rm p}=1.21)=-0.94\pm0.20$. The pivot redshift for the combination of all datasets is
$z_{\rm p}=0.23$, which corresponds to our tightest constraint on the dark energy equation of state, with 
$w_{\rm DE}(z_{\rm p}=0.23)=-1.040\pm0.072$, in good agreement with a cosmological constant.

\subsubsection{Dark energy and curvature}
\label{ssec:wok}

\begin{figure}
\centering
\centerline{\includegraphics[width=\columnwidth]{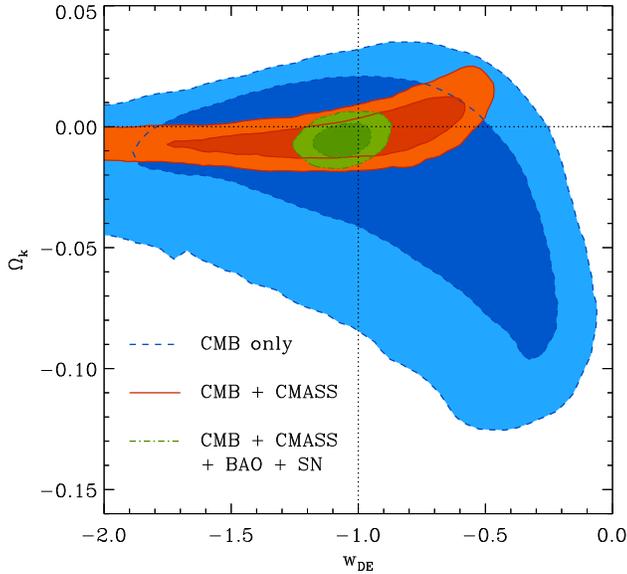}}
\caption{
The marginalized posterior distribution in the $w_{\rm DE}-\Omega_k$ plane for the $\Lambda$CDM 
parameter set extended by allowing for simultaneous variations on $w_{\rm DE}$ (assumed time-independent) and $\Omega_k$.
The dashed lines show the 68 and 95 per cent contours obtained using CMB information alone. The solid 
contours correspond to the results obtained from the combination of CMB data plus the shape of the CMASS $\xi(s)$.
The dot-dashed lines indicate the results obtained from the full dataset combination (CMB+CMASS+SN+BAO).
The dotted lines correspond to the values of these parameters in the $\Lambda$CDM model.
}
\label{fig:wok}
\end{figure}

We now explore the constraints on the dark energy equation of state (assumed time-independent) 
when the assumption of a flat Universe is dropped. 
This parameter space presents similar characteristics to the one studied in Section~\ref{ssec:wa},
where the dark energy equation of state is allowed evolve over time.
As discussed by \citet{Komatsu2009} and \citet{Sanchez2009},
when both $w_{\rm DE}$ and $\Omega_k$ are allowed to vary, the one-dimensional degeneracies
corresponding to constant values of $\Theta$ obtained from the CMB observations in the analyses of 
Sections~\ref{ssec:omk} and \ref{ssec:wde} gain an extra degree of freedom.
As shown by the dashed lines in Fig.~\ref{fig:wok},
when projected in the $w_{\rm DE}-\Omega_k$ plane, this two-dimensional degeneracy extends over a large 
region of the parameter space. 
The solid lines in Fig.~\ref{fig:wok} show the resulting constraints from the CMB+CMASS combination.
Although the constraint on $y_{\rm s}(z_{\rm m})$ provided by the CMASS correlation function 
substantially reduces the allowed region for these parameters, the remaining
degeneracy between them corresponds to poor one-dimensional marginalized restrictions.  
 
The distance measurements provided by the additional BAO or SN datasets can break the remaining
degeneracy, leading to meaningful constraints on these parameters. 
The dot-dashed lines in Fig.~\ref{fig:wok} correspond to the constraints obtained with 
the combination of all four datasets, 
showing good agreement with the $\Lambda$CDM model values (indicated by the dotted lines).
In this case, we obtain
$\Omega_k=-0.0054\pm0.0044$ and $ w_{\rm DE}=-1.060\pm0.075$, with similar accuracies 
to the constraints obtained when each of these parameters are varied independently (Sections~\ref{ssec:omk}
and \ref{ssec:wde}).

\begin{figure}
\centering
\centerline{\includegraphics[width=\columnwidth]{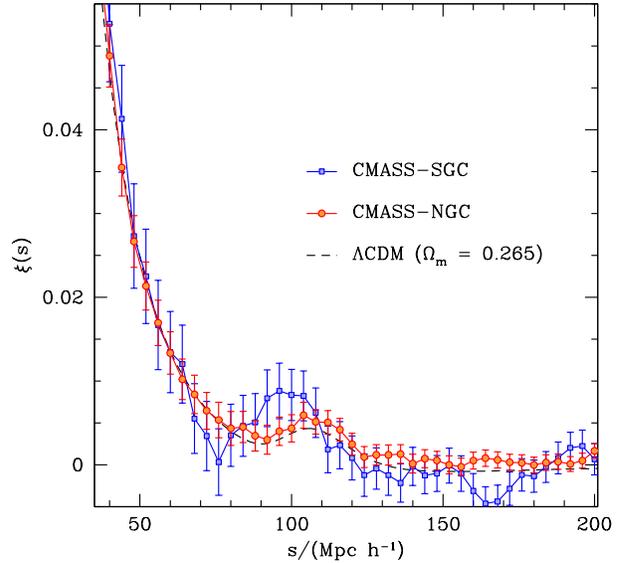}}
\caption{
Large-scale correlation function of the NGC (circles) and SGC (squares) CMASS sub-samples. The dashed line corresponds
to the best-fitting $\Lambda$CDM model obtained by combining the CMB data with the information from the shape of the NGC
correlation function. Although the two measurements exhibit the same broad-band shape, in the SGC data the BAO peak
has a larger amplitude and is located at smaller scales than in the NGC $\xi(s)$.
}
\label{fig:xi_ns}
\end{figure}

\section{The clustering signal in the Northern and Southern Galactic hemispheres}
\label{ssec:northsouth}

Our analysis is based on the full CMASS sample, combining the NGC and SGC data.
Compared to the NGC, the SGC observations correspond to a region with larger average Galactic extinction and
were taken under higher airmass and sky background and over different periods of time. 
These differences make the NGC-SGC split a sensible cut to study the clustering properties
of these sub-samples individually.
In fact, when analysed separately, the clustering of the NGC and SGC CMASS sub-samples
presents some intriguing differences. This can be seen in
Fig.~\ref{fig:xi_ns}, which shows the measurements of $\xi(s)$ in these two regions, obtained
as described in Section~\ref{sec:corfunc}. It is clear that, 
although they exhibit the same overall shape, the BAO feature in the SGC has a higher amplitude,
and its centroid is located at smaller scales than in the NGC.
In this Section, we explore the significance of these differences and their implications
on the obtained cosmological constraints.

\citet{Ross2012} performed a comprehensive analysis of the differences between the NGC and SGC CMASS sub-samples
and found no treatment of the data that could alleviate them.
\citet{Schlafly2010} and \citet{Schlafly2011} found small systematic offsets
between the colours of SDSS objects in the NGC and SGC, which lead to slightly different selection
criteria for the CMASS sample in the two galactic hemispheres.
\citet{Ross2011} found a 3.2\% difference in the number density of CMASS targets between
the NGC and SGC, which reduces to 0.3\% when the offset of \citet{Schlafly2011} is applied to the
galaxies in the SGC before applying the CMASS selection criteria.
 However, \citet{Ross2012} found that these factors do not produce a measurable effect on the clustering signal
of the SGC CMASS sample, and the differences between the correlation function of the SGC and NGC remain the same.

The consistency between the measurements in the NGC and SGC can be assessed by 
examining the difference $\xi_{\rm NGC}(s)-\xi_{\rm SGC}(s)$. As these regions
correspond to well separated volumes, we can neglect the covariance between them
and estimate the covariance matrix for this difference simply as 
$\mathbfss{ C}_{\rm diff}=\mathbfss{ C}_{\rm NGC}+\mathbfss{ C}_{\rm SGC}$,
where $\mathbfss{ C}_{\rm NGC}$ and $\mathbfss{ C}_{\rm SGC}$ correspond to the covariance matrices
of the individual NGC and SGC regions.
The consistency of the difference $\xi_{\rm NGC}(s)-\xi_{\rm NGC}(s)$ with cosmic variance
 can be estimated from its $\chi^2$ value, with respect to  $\mathbfss{ C}_{\rm diff}$.
In the range of scales used in our analysis, $40 < s/(h^{-1}{\rm Mpc}) < 200$,
we find $\chi^2=53.9$ for 41 data points. This number changes to 
$\chi^2=25.2$ for 15 points if the test is restricted to the range of scales of the BAO peak 
($70 < s/(h^{-1}{\rm Mpc}) < 130$). Using a different bin size of $\Delta s=7\,h^{-1}{\rm Mpc}$,
\citet{Ross2012} performed the same test and found similar values of $\chi^2$ per degree of freedom.
This result shows quantitatively that the general shapes of these measurements are in agreement, and the
differences between them are localized at the scales of the acoustic peak.
Note, however, that this is the range of scales from where the constraints on $y_{\rm s}$ are obtained.

\begin{figure}
\centering
\centerline{\includegraphics[width=\columnwidth]{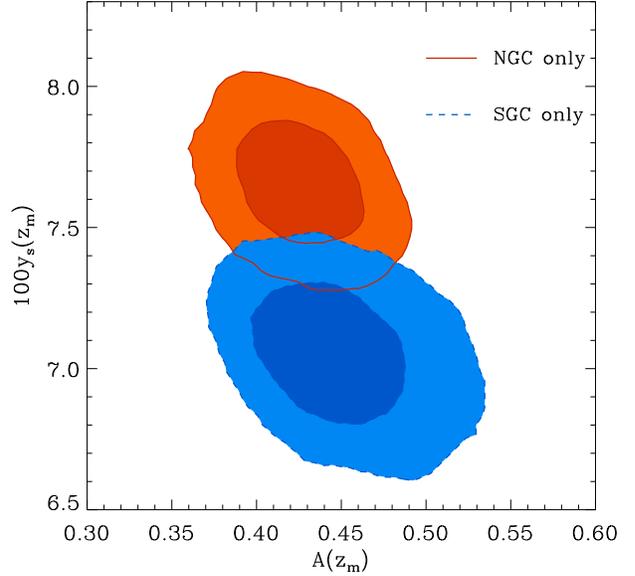}}
\caption{
The marginalized posterior distribution in the $A(z_{\rm m})-y_{\rm s}(z_{\rm m})$ plane obtained from the 
correlation function of the NGC (solid lines) and SGC (dashed lines) CMASS sub-samples. 
The contours correspond to the 68 and 95 per cent CL. 
While the two measurements point towards consistent values of $A(z_{\rm m})$,
their preferred values of $y_{\rm s}(z_{\rm m})$
deviate by approximately 2~$\sigma$.
}
\label{fig:tension_ns}
\end{figure}

Another view of this is presented in Fig.~\ref{fig:tension_ns}, which shows the two-dimensional 
marginalized constraints in the $A(z_{\rm m})-y_{\rm s}(z_{\rm m})$ plane. While the two measurements
point towards consistent values of $A(z_{\rm m})$, with $A(z_{\rm m})=0.426\pm0.021$ and $A(z_{\rm m})=0.447\pm0.030$
from the NGC and SGC, respectively, the different locations of the acoustic peak inferred from these regions
lead to $y_{\rm s}(z_{\rm m})=0.0762 \pm 0.0015$ and $y_{\rm s}(z_{\rm m})=0.0704 \pm 0.0017$,
which are approximately 2~$\sigma$ apart.
Despite the fact that the errors in the SCG correlation function are almost a factor two larger than those of its
NGC counterpart, the accuracies of the constraints on $y_{\rm s}$ obtained from these measurements
are similar. This is due to the high amplitude of the BAO bump in the SGC $\xi(s)$ which,
as can be seen in Fig.~\ref{fig:xi_ns}, gives a precise determination of the centroid of the peak, leading to
a slightly smaller than expected uncertainty on $y_{\rm s}$.
As was pointed out in Section~\ref{ssec:lcdm}, within the $\Lambda$CDM parameter space the CMB data alone
is sufficient to obtain the estimate $y_{\rm s}(z_{\rm m})=0.0762\pm0.0018$. This value shows a remarkable
consistency with the result obtained from the NGC. A comparison of Figs.~\ref{fig:corfunc} and \ref{fig:xi_ns}
shows that, although the correlation function of the full CMASS sample is dominated by that of the NGC, which
covers a larger volume, adding the data from the SGC moves the BAO peak towards somewhat smaller scales, leading
to the result $y_{\rm s}(z_{\rm m})=0.0742\pm0.0014$.

The conclusion from the tests of \citet{Ross2012} is that the differences between the NGC and SGC are simply due
to a statistical fluctuation. However, as the data in the NGC covers a volume 3.7~times larger, 
providing a better knowledge of the survey selection function,
for completeness we also discuss here the constraints on the 
parameter spaces of Section~\ref{ssec:param} obtained from the combination of the correlation function
of the NGC sub-sample with our CMB dataset. We do not consider here, however, the extension of the $\Lambda$CDM
parameter space in which both $w_{\rm DE}$ and $\Omega_k$ are allowed to float since, as discussed in Section~\ref{ssec:wok},
the combination of CMB data with a measurement of $\xi(s)$ is not enough to break the strong degeneracy
between these parameters.
The complete lists of parameter constraints obtained from the CMB+NGC combination is summarized in
Table~\ref{tab:ngc} of Appendix~\ref{sec:tables}. 

For the $\Lambda$CDM parameter space, the mean values for the cosmological parameters 
obtained in the CMB+NGC case are in closer agreement with those obtained by means of the CMB data alone than
in the full CMB+CMASS case. For example, in the CMB+NGC case we find constraints of
$\Omega_{\rm m}=0.265\pm0.014$ and $h=0.711\pm 0.012$, in excellent agreement with 
the CMB only results of $\Omega_{\rm m}=0.266\pm 0.024$ and $h=0.710\pm0.20$.
The slightly higher value of the Hubble parameter obtained in this case reduces the difference
with the measurement of \citet{Riess2011} to the 1~$\sigma$ level.

When the curvature of the Universe is included as a free parameter, the value of $y_{\rm s}(z_{\rm m})$ from
the NGC breaks the geometrical degeneracy in the CMB data closer to the locus of the flat models, yielding a
constraint of $\Omega_{k}=-0.0002\pm0.0049$, completely consistent with the flat Universe prediction
from the inflationary paradigm.

When constraining the fraction of massive neutrinos, the CMB+NGC combination yields
$f_{\nu}<0.044$ and $\sum m_{\nu}<0.52$~eV at 95\% CL. These limits are slightly tighter
than those obtained in the CMB+CMASS case. Regarding the constraints on the tensor-to-scalar ratio,
from the CMB+NGC combination we find $r < 0.17$ at the 95 per cent CL, which is equivalent to 
the limit found using the full CMASS $\xi(s)$, albeit with a preference for lower matter 
density values, with $\Omega_{\rm m}=0.276\pm0.016$

The results for the dark energy related parameter spaces also change when the full CMASS $\xi(s)$ is replaced
by the one of the NGC. In this case we obtain weaker constraints, with $\Omega_{\rm m}=0.246_{-0.42}^{0.045}$
and $w_{\rm DE}=-1.14\pm0.26$. When Equation~(\ref{eq:wa}) is used to explore the redshift dependence of the dark
energy equation of state, we find $w_{0}=-1.21_{-0.61}^{+0.79}$ and $w_{a}=0.1_{-1.0}^{+1.0}$ and a constraint of 
$w_{\rm DE}(z_{\rm p})=-1.21\pm0.26$ at the pivot redshift of $z_{\rm p}=0.96$. 

In all cases analysed, when we restrict our analysis to the NGC-CMASS sub-sample the constraints change at most by 1~$\sigma$.
This is in agreement with the results of \citet{Ross2012}, who found the same level of consistency.
In general, we find that the NGC data points towards slightly lower values of $\Omega_{\rm m}$ and higher ones of $h$
than those obtained from the full CMASS sample and in closer agreement with the CMB only case.
It should be emphasised, however, that the extensive tests of \citet{Ross2012}, together with our internal investigations,
show no reason for preferring the measurements from the NGC alone to the measurements from the full CMASS sample,  
which provides our best picture of the clustering of galaxies at $z\simeq0.57$.

\section{Conclusions}
\label{sec:conclusions}

We have presented an analysis of the cosmological implications of the monopole of the
redshift-space two-point correlation function, $\xi(s)$, measured from BOSS-DR9 CMASS sample.
The large volume and average number density of this sample make it ideally suited 
for large-scale structure analysis. 
The information contained in the full shape of the CMASS $\xi(s)$ allowed us to obtain 
accurate constraints of the parameters $y_{\rm s}(z_{\rm m})$ and $A(z_{\rm m})$, given by equations~(\ref{eq:ys})
and (\ref{eq:apar}). 
By adopting an explicit, perturbation-theory based model for the correlation function in the mildly non-linear regime, and marginalizing
over its uncertain parameters, we are able to exploit information beyond that in the scale of the BAO peak alone.
We combined this information with that of additional cosmological probes,
including CMB, SN, and BAO measurements from other data sets, to derive constraints on cosmological parameters.
We studied the parameters of the $\Lambda$CDM parameter space, and a number of its extensions.
The main results from our analysis can be summarized as follows:

\begin{enumerate}

 \item Our results show that 
the simple $\Lambda$CDM model is able to describe all the datasets that we have included in our
analysis. Given the different nature of these observations and the range of redshifts they probe,
this is not a minor achievement. The basic parameters of this model are constrained to
an accuracy better than 5\%; a clear demonstration of the constraining power of
observations in the current era of precision cosmology. 

 \item Fig.~\ref{fig:extensions} summarizes our constraints on possible extensions of the standard $\Lambda$CDM
model. We considered non-flat models, massive neutrinos, non-zero primordial tensor fluctuations,
and more general dark energy models. 
In all of these cases the inclusion of the CMASS $\xi(s)$ in the analysis significantly improves the obtained
constraints with respect to those obtained using the CMB data alone.
Our results show no significant evidence of deviations from the
$\Lambda$CDM picture, which can  still be considered as our best cosmological model.

 \item The information provided by the CMASS correlation function is essential to obtain
tight constraints on the curvature of the Universe. We obtain the constraint $\Omega_{k}=-0.0043_{-0.0049}^{+0.0049}$
from the CMB+CMASS combination which is not significantly improved by adding information from SN or other BAO data.

 \item When massive neutrinos are considered in the analysis, we find a constraint of $f_{\nu}<0.056$ at the
95\% CL, implying a limit of $\sum m_{\nu} < 0.61\,{\rm eV}$ on the sum of the three neutrino species. This limit
is improved to $\sum m_{\nu} < 0.51 {\rm eV}$ when the SN and BAO data are added to the analysis.

 \item When considering tensor modes the CMB+CMASS combination allowed us to obtain a limit on the
tensor-to-scalar amplitude ratio of $r<0.16$ at the 95\% CL, which is almost unchanged by considering 
additional datasets. The combination of CMB data with the shape of the CMASS correlation function reveals 
a clear signature of a deviation from scale-invariance, with $n_{\rm s}<1$ at the 99.7\% CL, 
also in the presence of tensor modes.

  \item We explored models where the dark energy component does not correspond to a cosmological constant
and found no signature of a deviation from the standard $\Lambda$CDM model.
When the value of $w_{\rm DE}$, assumed time-independent, is allowed to vary, the CMB+CMASS combination provides the constraint
$w_{\rm DE}=-0.95_{-0.20}^{+0.21}$.
Interestingly, the CMASS+SN combination alone provides a tighter constraint, with $w_{\rm DE}=-1.04\pm0.11$, independently of any CMB data.
Our tighter constraints are obtained from the combination of all datasets, with $w_{\rm DE}=-1.033_{-0.074}^{+0.073}$,
in good agreement with a cosmological constant. This result does not change significantly if the assumption of a flat universe is
relaxed. We also find no evidence of a redshift evolution of $w_{\rm DE}$.

  \item Our results are in excellent agreement with those of \citet{Blanton2012} and \citet{Reid2012}, who explored 
the cosmological implications of the BAO and redshift-space distortions measurements in the CMASS sample.
This highlights the consistency between the different analysis techniques implemented in each of these studies,
and provides a reassuring demonstration of the robustness of our results. 

  \item We studied the clustering of the NGC and SGC regions separately. The overall shapes of the 
correlation functions in these two sub-samples show good agreement, but they differ in the location 
and amplitude of the BAO peak. This translates into constraints of $y_{\rm s}(z_{\rm m})$
which differ at the 2~$\sigma$ level. \citet{Ross2012} performed a detailed analysis of the clustering signal
in these regions and found no evidence of additional systematic effects in the SGC data, indicating that the observed
differences are simply due to a statistical fluctuation. For completeness, we explored the constraints obtained
when the NGC correlation function is used in combination with our CMB datasets. In all cases our results remain
unchanged within 1~$\sigma$, with the NGC data pointing towards slightly lower values of $\Omega_{\rm m}$ and
higher ones of $h$ than those obtained from the full CMASS sample.

\end{enumerate}

The current analysis is based on the first spectroscopic data release of BOSS. The larger volume that will be
probed by subsequent data releases, plus the extension of the analysis to the lower redshift BOSS galaxies,
will reduce the uncertainties in the measurement of $\xi(s)$ and the calibration of the corrections for
potential systematic effects, providing even more accurate views of the LSS clustering pattern in the Universe.
This improvement will be accompanied by the release of the CMB power spectra measurement from the Planck satellite
in early 2013. The combination of these datasets will undoubtedly provide new, more stringent constraints on
cosmological parameters, and open up the possibility to explore additional extensions to the $\Lambda$CDM 
model which have not yet been fully explored.

\section*{Acknowledgements}

We would like to thank Ryan Keisler for his help with the implementation of the SPT likelihood code in {\sc CosmoMC}. 
We also thank Bradford Benson for helping us find a bug in our modifications to {\sc CosmoMC} by pointing out a discrepancy between our
constraints on the equation of state of dark energy using only CMB data and the reported values by the WMAP team.
AGS would like to thank all users of the Pan-STARRS cluster in Garching for their patience and support.
CGS and JAR-M acknowledge funding from project AYA2010-21766-C03-02 of the Spanish Ministry of Science and
Innovation (MICINN). 
AJR is grateful to the UK Science and Technology Facilities Council
for financial support through the grant ST/I001204/1. WJP is
grateful for support from the the UK Science and Technology Facilities
Research Council, the Leverhulme Trust, and the European Research Council.
FP acknowledges support from the Spanish MICINN’s Consolider grant MultiDark CSD2009-00064.
JAR-M is a Ram\'on y Cajal fellow of the Spanish Ministry of Science and Innovation (MICINN).

Funding for SDSS-III has been provided by the Alfred P. Sloan Foundation, the Participating Institutions, 
the National Science Foundation, and the U.S. Department of Energy. 

SDSS-III is managed by the Astrophysical Research Consortium for the
Participating Institutions of the SDSS-III Collaboration including the
University of Arizona,
the Brazilian Participation Group,
Brookhaven National Laboratory,
University of Cambridge,
Carnegie Mellon University,
University of Florida,
the French Participation Group,
the German Participation Group,
Harvard University,
the Instituto de Astrofisica de Canarias,
the Michigan State/Notre Dame/JINA Participation Group,
Johns Hopkins University,
Lawrence Berkeley National Laboratory,
Max Planck Institute for Astrophysics,
Max Planck Institute for Extraterrestrial Physics,
New Mexico State University,
New York University,
Ohio State University,
Pennsylvania State University,
University of Portsmouth,
Princeton University,
the Spanish Participation Group,
University of Tokyo,
University of Utah,
Vanderbilt University,
University of Virginia,
University of Washington,
and Yale University.

We acknowledge the use of the Legacy Archive for Microwave Background 
Data Analysis (LAMBDA). Support for LAMBDA is provided by the NASA
Office of Space Science. 


\appendix

\section{Summary of the obtained cosmological constraints} 
\label{sec:tables}

In this section we summarize the constraints on cosmological parameters obtained
using different combinations of the datasets described in Sections~\ref{sec:corfunc} and \ref{sec:moredata}.
Table~\ref{tab:lcdm} lists the 68\% confidence limits on the parameters of the $\Lambda$CDM model, as discussed in Section~\ref{ssec:lcdm}.
Tables~\ref{tab:omk}-\ref{tab:wok} correspond to the extensions of this parameter space analysed in Sections~\ref{ssec:omk} to \ref{ssec:darkenergy}.
Finally, Table~\ref{tab:ngc} presents the constraints on these parameter spaces, obtained from the combination of the correlation function
of the NGC sub-sample with the CMB data.

\begin{table*} 
\centering
  \caption{
    The marginalized 68\% constraints on the cosmological parameters of the $\Lambda$CDM model
    obtained using different combinations of the datasets described in Section~\ref{ssec:cmass}
    and \ref{sec:moredata}.}
    \begin{tabular}{@{}lccccc@{}}
    \hline
& \multirow{2}{*}{CMB}  & \multirow{2}{*}{CMB + CMASS} & CMB + CMASS & CMB + CMASS & CMB + CMASS \\
&                       &                           &  +SN     &    +BAO  &   + BAO + SN       \\  
\hline
$100\,\Theta$           & $1.0411_{-0.0016}^{+0.0016}$  & $1.0407_{-0.0015}^{+0.0015}$  & $1.0408_{-0.0015}^{+0.0015}$ &  $1.0406_{-0.0015}^{+0.0015}$  &  $1.0406_{   -0.0015}^{+0.0015}$   \\[1.5mm] 
$100\,\omega_{\rm b}$   & $2.223_{-0.041}^{+0.041}$   & $2.21_{-0.039}^{+0.039}$    & $2.22_{-0.039}^{+0.039}$       &  $2.21_{-0.038}^{+0.038}$      &  $2.21_{-0.038}^{+0.038}$ \\[1.5mm] 
$100\,\omega_{\rm c}$   & $11.16_{-0.45}^{+0.45}$     & $11.45_{-0.29}^{+0.28}$  & $11.35_{-0.28}^{+0.28}$ &  $11.58_{-0.22}^{+0.22}$  &  $11.50_{-0.20}^{+0.20}$ \\[1.5mm] 
$\tau $                 & $0.0857_{-0.0068}^{+0.0061}$  & $0.0822_{-0.0064}^{+0.0060}$  & $0.0834_{-0.0068}^{+0.0059}$ &  $0.0811_{-0.0062}^{+0.0056}$  &  $0.0815_{-0.0065}^{+0.0059}$   \\[1.5mm] 
$n_{\rm s}$             & $0.967_{-0.011}^{+0.010}$  & $0.9620_{-0.0091}^{+0.0093}$  & $0.9638_{-0.0092}^{+0.0091}$ &  $ 0.9604_{-0.0087}^{+0.0087}$  &  $0.9613_{-0.0090}^{+0.0089}$   \\[1.5mm] 
$\ln(10^{10}A_{\rm s})$ & $3.082_{-0.030}^{+0.030}$  & $3.085_{-0.028}^{+0.028}$  & $3.084_{-0.029}^{+0.029}$ &  $3.086_{-0.027}^{+0.028}$  &  $3.084_{-0.028}^{+0.028}$ \\[1.5mm] 
$\Omega_{\rm DE}$       & $0.734_{-0.024}^{+0.024}$  & $0.718_{-0.015}^{+0.015}$  & $0.724_{-0.014}^{+0.014}$ &  $0.711_{-0.010}^{+ 0.010}$  &  $0.7154_{-0.0094}^{+0.0097}$  \\[1.5mm]
$\Omega_{\rm m}$        & $0.266_{-0.024}^{+0.024}$  & $0.282_{-0.015}^{+0.015}$  & $0.276_{-0.014}^{+0.014}$ &  $0.289_{-0.010}^{+0.010}$  &  $0.2846_{-0.0097}^{+0.0095}$  \\[1.5mm]
$\sigma_{8}$            & $0.814_{-0.023}^{+0.023}$  & $0.825_{-0.018}^{+0.018}$  & $0.821_{-0.018}^{+0.018}$ &  $0.830_{-0.016}^{+0.016}$  &  $0.827_{-0.016}^{+0.016}$    \\[1.5mm]
$t_{0}/{\rm Gyr}$       & $13.725_{-0.084}^{+0.086}$  & $13.769_{-0.071}^{+0.072}$  & $13.753_{-0.072}^{+0.072}$ &  $13.780_{-0.066}^{+0.066}$  &  $13.774_{-0.068}^{+0.067}$   \\[1.5mm] 
$z_{\rm re}$            & $10.4_{-1.2}^{+1.2}$       & $10.2_{-1.1}^{+1.2}$       & $10.3_{-1.2}^{+1.1}$      &  $10.2_{-1.1}^{+1.1}$       &  $10.2_{-1.2}^{+1.2}$ \\[1.5mm]
$h$                     & $0.710_{-0.020}^{+0.020}$  & $0.696_{-0.012}^{+0.012}$  & $0.701_{-0.012}^{+0.012}$ &  $0.691_{-0.084}^{+0.084}$  &  $0.694_{-0.081}^{+0.082}$ \\[1.5mm]
$D_{\rm V}(z_{\rm m})/{\rm Mpc}$  & $2006_{-32}^{+33}$  & $ 2028_{-20}^{+20}$         & $2020_{-20}^{+20}$        &  $ 2036_{-15}^{+15}$          &  $ 2031_{-15}^{+15}$    \\[1.5mm]
$f(z_{\rm m})$          & $0.743_{-0.021}^{+0.021}$  & $0.757_{-0.012}^{+0.012}$  & $0.752_{-0.012}^{+0.012}$ &  $0.7628_{-0.0083}^{+0.0082}$  &  $0.7595_{-0.0078}^{+0.0077}$  \\
\hline
\end{tabular}
\label{tab:lcdm}
\end{table*}

\begin{table*} 
\centering
  \caption{
    The marginalized 68\% allowed regions on the cosmological parameters of the $\Lambda$CDM model extended by adding $\Omega_k$ as a free parameter,
    obtained using different combinations of the datasets described in Section~\ref{ssec:cmass}
    and \ref{sec:moredata}.}
    \begin{tabular}{@{}lccccc@{}}
    \hline
& \multirow{2}{*}{CMB}  & \multirow{2}{*}{CMB + CMASS} & CMB + CMASS & CMB + CMASS & CMB + CMASS \\
&                       &                           &  +SN     &    +BAO  &   + BAO + SN       \\  
\hline
$\Omega_K$  &  $ -0.014_{-0.025}^{+0.022}$ &  $-0.0042_{-0.0049}^{+0.0050}$ &  $-0.0047_{ -0.0048}^{+0.0047}$ &  $-0.0042_{-0.0043}^{+ 0.0044}$ &  $-0.0045_{-0.0042}^{+0.0043}$ \\[1.5mm]
100$\Theta$  &  $1.0411_{-0.0016}^{+0.0016}$ &  $1.0411_{-0.0016}^{+0.0016}$ &  $1.0411_{-0.0016}^{+0.0016}$ &  $ 1.0411_{   -0.0016}^{+ 0.0016}$ &  $1.0410_{-0.0015}^{+0.0015}$ \\[1.5mm]
100$\omega_b$  &  $2.221_{-0.041}^{+0.043}$ &  $ 2.220_{-0.040}^{+0.040}$ &  $2.227_{-0.040}^{+    0.041}$ &  $ 2.222_{-0.041}^{+ 0.040}$ &  $ 2.223_{-0.037}^{+ 0.039}$ \\[1.5mm]
100$\omega_{\rm dm}$  &  $ 11.20_{-0.47}^{+0.46}$ &  $ 11.19_{-0.43}^{+ 0.44}$ &  $ 11.04_{   -0.42}^{+ 0.41}$ &  $ 11.24_{-0.41}^{+    0.42}$ &  $    11.13_{   -0.40}^{+ 0.40}$ \\[1.5mm]
$\tau$  &  $ 0.0840_{ -0.0071}^{+0.0062}$ &  $ 0.0842_{-0.0066}^{+0.0058}$ &  $0.0862_{-0.0068}^{+ 0.0060}$ &  $ 0.0850_{-0.0071}^{+0.0064}$ &  $ 0.0848_{-0.0074}^{+ 0.0064}$ \\[1.5mm]
$n_{\rm s}$  &  $   0.965_{-0.011}^{+0.011}$ &  $ 0.965_{-0.010}^{+ 0.010}$ &  $ 0.967_{-0.0098}^{+ 0.0099}$ &  $ 0.965_{ -0.010}^{+ 0.0010}$ &  $ 0.966_{-0.0095}^{+ 0.0098}$ \\[1.5mm]
ln($10^{10}A_{\rm s}$)  &  $ 3.079_{ -0.030}^{+0.029}$ &  $ 3.079_{- 0.030}^{+    0.030}$ &  $    3.078_{-0.030}^{+    0.029}$ &  $    3.083_{  -0.030}^{+    0.031}$ &  $    3.078_{   -0.031}^{+    0.032}$ \\[1.5mm]
$\Omega_{\rm DE}$  &  $  0.693_{   -0.079}^{+0.074}$ &  $    0.719_{-0.015}^{+    0.016}$ &  $    0.726_{   -0.014}^{+    0.014}$ &  $    0.717_{   -0.012}^{+    0.012}$ &  $    0.721_{   -0.012}^{+    0.012}$ \\[1.5mm]
$\Omega_{\rm m}$  &  $    0.321_{   -0.094}^{+    0.104}$ &  $ 0.285_{   -0.016}^{+ 0.015}$ &  $    0.279_{   -0.015}^{+    0.015}$ &  $    0.287_{ -0.010}^{+    0.011}$ &  $    0.283_{-0.010}^{+    0.010}$ \\[1.5mm]
$\sigma_{\rm 8}$  &  $ 0.806_{   -0.027}^{+ 0.027}$ &  $    0.812_{   -0.024}^{+    0.024}$ &  $    0.806_{   -0.023}^{+    0.023}$ &  $    0.815_{   -0.023}^{+    0.023}$ &  $    0.809_{   -0.023}^{+    0.024}$ \\[1.5mm]
$t_{\rm 0}/{\rm Gyr}$  &  $   14.20_{  -1.00}^{+  1.07}$ &  $   13.95_{   -0.23}^{+0.22}$ &  $   13.96_{   -0.21}^{+    0.23}$ &  $   13.95_{   -0.20}^{+ 0.20}$ &  $13.97_{ -0.20}^{+    0.19}$ \\[1.5mm]
$z_{\rm re}$  &  $   10.3_{   -1.2}^{+ 1.2}$ &  $   10.3_{   -1.2}^{+    1.2}$ &  $   10.4_{   -1.2}^{+    1.1}$ &  $   10.4_{   -1.2}^{+    1.2}$ &  $   10.3_{   -1.2}^{+    1.2}$ \\[1.5mm]
$h$  &  $   0.669_{  -0.106}^{+    0.097}$ &  $   0.687_{   -0.017}^{+    0.017}$ &  $   0.690_{   -0.016}^{+    0.016}$ &  $   0.685_{-0.011}^{+    0.011}$ &  $   0.687_{   -0.010}^{+    0.011}$ \\[1.5mm]
$D_{\rm V}(z_{\rm m})$  &  $ 2116_{ -222}^{+  242}$ &  $ 2057_{  -39}^{+ 39}$ &  $ 2053_{  -38}^{+   39}$ &  $ 2059_{  -29}^{+   29}$ &  $ 2057_{  -30}^{+   29}$ \\[1.5mm]
$f(z_{\rm m})$  &  $    0.779_{   -0.076}^{+    0.083}$ &  $    0.761_{ -0.013}^{+    0.013}$ &  $    0.756_{   -0.013}^{+    0.013}$ &  $    0.7629_{   -0.0085}^{+    0.0085}$ &  $    0.7600_{ -0.0086}^{+    0.0083}$ \\[1.5mm]
\hline
\end{tabular}
\label{tab:omk}
\end{table*}

\begin{table*} 
\centering
  \caption{
    The marginalized 68\% allowed regions on the cosmological parameters of the $\Lambda$CDM model extended by adding $f_{\nu}$ as a free parameter,
    obtained using different combinations of the datasets described in Section~\ref{ssec:cmass}
    and \ref{sec:moredata}.}
    \begin{tabular}{@{}lccccc@{}}
    \hline
& \multirow{2}{*}{CMB}  & \multirow{2}{*}{CMB + CMASS} & CMB + CMASS & CMB + CMASS & CMB + CMASS \\
&                       &                           &  +SN     &    +BAO  &   + BAO + SN       \\  
\hline
$f_\nu$  &  $< 0.11 $ (95\% CL) & $< 0.055$ (95\% CL)  & $ <0.049$ (95\% CL)  & $ < 0.050 $ (95\% CL) &  $ < 0.049 $ (95\% CL) \\[1.5mm] 
100$\Theta$  &  $    1.0405_{ -0.0016}^{+    0.0016}$ &  $    1.0407_{   -0.0015}^{+    0.0015}$ &  $    1.0408_{   -0.0014}^{+    0.0014}$ &  $    1.0408_{   -0.0015}^{+    0.0015}$ &  $    1.0409_{   -0.0015}^{+    0.0014}$ \\[1.5mm]
100$\omega_{\rm b}$  &  $    2.191_{   -0.047}^{+ 0.046}$ &  $ 2.214_{   -0.040}^{+ 0.040}$ &  $ 2.219_{   -0.039}^{+ 0.038}$ &  $    2.213_{   -0.038}^{+  0.038}$ &  $ 2.217_{ -0.039}^{+    0.039}$ \\[1.5mm]
100$\omega_{\rm dm}$  &  $  12.12_{   -0.78}^{+    0.79}$ &  $    11.53_{   -0.29}^{+    0.29}$ &  $  11.38_{   -0.27}^{+    0.28}$ &  $ 11.5076_{   -0.21}^{+    0.20}$ &  $  11.45_{ -0.21}^{+    0.21}$ \\[1.5mm]
$\tau$  &  $ 0.0829_{-0.0066}^{+    0.0060}$ &  $    0.0852_{ -0.0067}^{+ 0.0059}$ &  $ 0.0860_{ -0.0064}^{+ 0.0057}$ &  $    0.0845_{ -0.0069}^{+0.0064}$ &  $0.0856_{-0.0074}^{+ 0.0062}$ \\[1.5mm]
$n_{\rm s}$  &  $ 0.956_{-0.014}^{+ 0.014}$ &  $ 0.965_{-0.009}^{+ 0.010}$ &  $ 0.966_{-0.009}^{+ 0.009}$ &  $0.964_{ -0.009}^{+ 0.009}$ &  $ 0.966_{-0.010}^{+ 0.010}$ \\[1.5mm]
ln$(10^{10} A_{\rm s})$  &  $ 3.079_{ -0.029}^{+ 0.029}$ &  $ 3.082_{ -0.028}^{+ 0.029}$ &  $  3.080_{ -0.028}^{+ 0.029}$ &  $ 3.080_{   -0.029}^{+ 0.030}$ &  $  3.082_{-0.030}^{+ 0.031}$ \\[1.5mm]
$\sum m_\nu$  &  $ < 1.4\,{\rm eV} $ (95\% CL) & $ < 0.61\,{\rm eV} $ (95\% CL)  & $ < 0.52\,{\rm eV} $ (95\% CL)  & $ < 0.54\,{\rm eV} $ (95\% CL) &  $ < 0.51\,{\rm eV}$ (95\% CL) \\[1.5mm] 
$\Omega_{\rm DE}$  &  $    0.643_{ -0.073}^{+ 0.070}$ &  $ 0.702_{  -0.020}^{+ 0.020}$ &  $ 0.712_{ -0.016}^{+ 0.016}$ &  $ 0.704_{  -0.011}^{+ 0.011}$ &  $  0.708_{-0.011}^{+ 0.011}$ \\[1.5mm]
$\Omega_{\rm m}$  &  $ 0.357_{ -0.070}^{+ 0.073}$ &  $  0.298_{ -0.019}^{+ 0.019}$ &  $ 0.288_{ -0.016}^{+ 0.016}$ &  $ 0.296_{-0.010}^{+ 0.011}$ &  $ 0.292_{ -0.011}^{+  0.011}$ \\[1.5mm]
$\sigma_{\rm 8}$  &  $    0.683_{ -0.079}^{+ 0.081}$ &  $ 0.752_{ -0.048}^{+0.484}$ &  $ 0.759_{ -0.045}^{+ 0.046}$ &  $ 0.756_{   -0.049}^{+    0.051}$ &  $  0.758_{ -0.046}^{+ 0.046}$ \\[1.5mm]
$t_0/$Gyr  &  $   14.116_{ -0.258}^{+ 0.251}$ &  $ 13.902_{-0.112}^{+ 0.110}$ &  $   13.865_{   -0.097}^{+    0.099}$ &  $   13.890_{ -0.093}^{+ 0.094}$ &  $   13.873_{-0.089}^{+ 0.088}$ \\[1.5mm]
$z_{\rm re}$  &  $   10.507_{   -1.145}^{+    1.194}$ &  $   10.519_{   -1.135}^{+    1.172}$ &  $   10.535_{   -1.136}^{+    1.137}$ &  $   10.452_{   -1.207}^{+    1.139}$ &  $   10.523_{   -1.242}^{+    1.283}$ \\[1.5mm]
$h$  &  $   0.640_{   -0.048}^{+    0.048}$ &  $   0.680_{   -0.0160}^{+ 0.016}$ &  $   0.688_{   -0.014}^{+    0.014}$ &  $   0.681_{  -0.011}^{+ 0.011}$ &  $   0.685_{ -0.010}^{+ 0.010}$ \\[1.5mm]
$D_{\rm V}(z_m)$  &  $ 2135_{  -87}^{+   87}$ &  $ 2062_{  -29}^{+   30}$ &  $ 2047_{  -26}^{+   25}$ &  $ 2058_{  -20}^{+   21}$ &  $ 2052_{  -20}^{+   20}$  \\[1.5mm]
$f(z_{\rm m})$  &  $    0.806_{ -0.044}^{+    0.045}$ &  $  0.770_{   -0.014}^{+    0.014}$ &  $    0.762_{   -0.013}^{+    0.013}$ &  $    0.768_{   -0.008}^{+    0.009}$ &  $    0.765_{   -0.009}^{+    0.0089}$ \\
\hline
\end{tabular}
\label{tab:fnu}
\end{table*}

\begin{table*}
\centering
  \caption{
    The marginalized 68\% allowed regions on the cosmological parameters of the $\Lambda$CDM model extended by allowing for non-zero primordial tensor modes,
    obtained using different combinations of the datasets described in Section~\ref{ssec:cmass}
    and \ref{sec:moredata}.}
    \begin{tabular}{@{}lccccc@{}}
    \hline
& \multirow{2}{*}{CMB}  & \multirow{2}{*}{CMB + CMASS} & CMB + CMASS & CMB + CMASS & CMB + CMASS \\
&                       &                           &  +SN     &    +BAO  &   + BAO + SN       \\
\hline
$r$  &  $<0.21$  (95\% CL) & $<0.16$ (95\% CL) & $<0.16$  (95\% CL)& $<0.15$ (95\% CL)& $<0.15$  (95\% CL) \\[1.5mm]
100$\Theta$  &  $1.0413_{-0.0016}^{+0.0016}$ & $1.0408_{-0.0015}^{+0.0015}$ &  $1.0409_{-0.0015}^{+0.0015}$ & $1.0406_{-0.0015}^{+0.0015}$ &  $1.0407_{-0.0015}^{+0.0015}$ \\[1.5mm]
100$\omega_{\rm b}$  & $2.240_{-0.045}^{+0.045}$ &  $2.221_{-0.040}^{+0.040}$ &  $2.228_{-0.038}^{+0.038}$ & $2.215_{-0.038}^{+0.039}$ &  $2.219_{-0.038}^{+0.039}$\\[1.5mm]
100$\omega_{\rm dm}$  & $ 10.95_{-0.48}^{+0.50}$ &  $11.42_{-0.31}^{+0.31}$ &  $11.31_{-0.28}^{+0.27}$ &  $ 11.55_{-0.21}^{+0.21}$ &  $11.47_{-0.20}^{+0.20}$\\[1.5mm]
$\tau$  & $0.0856_{-0.0071}^{+0.0062}$ &  $0.0815_{-0.0067}^{+0.0059}$ &  $0.0825_{-0.0068}^{+0.0061}$ &  $0.0808_{-0.0064}^{+0.0060}$ &  $    0.0812_{-0.0064}^{+0.0060}$\\[1.5mm]
$n_{\rm s}$  & $0.974_{-0.013}^{+0.013}$ & $0.966_{-0.011}^{+0.010}$ &  $ 0.9679_{-0.0096}^{+0.0094}$ &  $ 0.9636_{-0.0096}^{+0.0094}$ &  $    0.9652_{-0.0093}^{+0.0093}$\\[1.5mm]
ln$(10^{10} A_{\rm s})$  & $3.077_{-0.029}^{+0.030}$ & $3.083_{-0.028}^{+0.028}$ &  $3.082_{-0.029}^{+0.030}$ &  $3.086_{-0.028}^{+0.028}$ &  $3.084_{-0.029}^{+0.029}$\\[1.5mm]
$\Omega_{\rm DE}$  & $0.745_{-0.025}^{+0.025}$ &  $0.720_{-0.016}^{+0.016}$ &  $ 0.726_{-0.014}^{+0.014}$ &  $0.713_{-0.010}^{+0.010}$ &  $    0.7173_{-0.0098}^{+0.0098}$\\[1.5mm]
$\Omega_{\rm m}$  &$0.255_{-0.025}^{+0.025}$ & $0.280_{-0.016}^{+0.016}$ & $0.274_{-0.014}^{+0.014}$ &  $0.287_{-0.010}^{+0.010}$ &  $    0.2827_{-0.0010}^{+0.0098}$\\[1.5mm]
$\sigma_{\rm 8}$  &$0.805_{-0.024}^{+0.025}$ &  $ 0.824_{-0.018}^{+0.018}$ & $0.820_{-0.018}^{+0.018}$ &  $0.830_{-0.016}^{+0.016}$ &  $    0.827_{-0.016}^{+0.016}$\\[1.5mm]
$t_0/$Gyr  &$13.690_{-0.095}^{+0.094}$ & $13.754_{-0.075}^{+0.075}$ & $13.738_{-0.071}^{+0.071}$ & $13.771_{-0.067}^{+0.068}$ &  $   13.763_{-0.069}^{+0.065}$ \\[1.5mm]
$z_{\rm re}$  &$10.3_{-1.1}^{+1.2}$ & $10.1_{-1.1}^{+1.2}$ & $10.2_{-1.2}^{+1.2}$ & $10.1_{-1.1}^{+1.2}$ & $10.1_{-1.2}^{+1.2}$\\[1.5mm]
$h$  & $0.721_{-0.023}^{+0.023}$ & $ 0.699_{-0.013}^{+0.013}$ &  $ 0.704_{-0.012}^{+0.012}$ & $ 0.6930_{-0.0083}^{+0.0085}$ &  $   0.6962_{-0.0083}^{+0.0084}$\\[1.5mm]
$D_{\rm V}(z_m)$  & $ 1990_{-37}^{+37}$ & $2024_{-22}^{+22}$ & $ 2016_{-20}^{+20}$ & $ 2033_{-15}^{+15}$ & $ 2028_{-15}^{+15}$ \\[1.5mm]
$f(z_{\rm m})$ & $ 0.733_{-0.023}^{+0.023}$ & $0.755_{-0.013}^{+0.013}$ &  $0.750_{-0.012}^{+0.012}$ &  $0.7613_{-0.0081}^{+0.0082}$ &  $    0.7580_{-0.0080}^{+0.0080}$  \\
\hline
\end{tabular}
\label{tab:r}
\end{table*}

\begin{table*}
\centering
  \caption{
    The marginalized 68\% allowed regions on the cosmological parameters of the $\Lambda$CDM model extended by including the redshift-independent value of $w_{\rm DE}$ as an additional parameter,
    obtained using different combinations of the datasets described in Section~\ref{ssec:cmass}
    and \ref{sec:moredata}.}
    \begin{tabular}{@{}lccccc@{}}
    \hline
& \multirow{2}{*}{CMB}  & \multirow{2}{*}{CMB + BOSS} & CMB + CMASS & CMB + CMASS & CMB + CMASS \\
&                       &                           &  +SN     &    +BAO  &   + BAO + SN       \\
\hline
$w_{\rm DE}$            &$-1.15_{-0.39}^{+0.39}$      & $-0.95_{-0.20}^{+0.21}$     & $-1.054_{-0.076}^{+0.077}$   & $-0.91_{-0.11}^{+0.11}$      & $-1.033_{-0.074}^{+0.073}$ \\[1.5mm]
$100\,\Theta$           &$1.0410_{-0.0016}^{+0.0016}$ & $1.0410_{0.0016}^{+0.0016}$ & $1.0406_{-0.0015}^{+0.0015}$ & $1.0409_{-0.00156}^{+0.0016}$& $1.0405_{-0.0015}^{+0.0015}$ \\[1.5mm]
$100\,\omega_{\rm b}$   &$2.219_{-0.042}^{+0.042}$    & $2.220_{-0.042}^{+0.043}$   & $2.211_{-0.038}^{+0.039}$    & $2.221_{-0.041}^{+0.041}$    & $2.210_{-0.039}^{+0.039}$  \\[1.5mm]
$100\,\omega_{\rm dm}$   & $11.21_{-0.47}^{+0.47}$      & $11.33_{-0.47}^{+0.48}$     & $11.48_{-0.33}^{+0.33}$      & $11.24_{-0.43}^{+0.43}$      & $11.58_{-0.32}^{+0.32}$  \\[1.5mm]
$\tau $                 & $0.0847_{-0.0071}^{+0.0060}$ & $0.0831_{-0.0070}^{+0.0063}$& $0.0819_{-0.0064}^{+0.0059}$ & $0.0840_{-0.0070}^{+0.0062}$ & $0.0814_{0.0063}^{+0.0057}$  \\[1.5mm]
$n_{\rm s}$             &$0.965_{-0.011}^{+0.011}$    & $0.964_{-0.011}^{+0.011}$   & $0.9615_{-0.0098}^{+0.0097}$ & $0.966_{-0.011}^{+0.011}$    & $0.9606_{-0.0095}^{+0.0096}$  \\[1.5mm]
$\ln(10^{10}A_{\rm s})$ &$3.081_{-0.030}^{+0.030}$    & $3.083_{-0.030}^{+0.030}$   & $3.084_{-0.028}^{+0.028}$    & $3.081_{-0.030}^{+0.030}$    & $3.087_{-0.028}^{+0.028}$  \\[1.5mm]
$\Omega_{\rm DE}$       &$0.751_{-0.093}^{+0.088}$    & $0.704_{-0.041}^{+0.042}$   & $0.729_{-0.016}^{+0.016}$    & $0.702_{-0.017}^{+0.017}$    & $0.719_{-0.012}^{+0.012}$  \\[1.5mm]
$\Omega_{\rm m}$        &$0.248_{-0.088}^{+0.093}$    & $0.295_{-0.042}^{+0.041}$   & $0.270_{-0.016}^{+0.016}$    & $0.297_{-0.017}^{+0.017}$    & $0.281_{-0.012}^{+0.012}$  \\[1.5mm]
$\sigma_{8}$            &$0.86_{-0.13}^{+0.13}$       & $0.801_{0.084}^{+0.085}$    & $0.842_{-0.035}^{+0.035}$    & $0.787_{-0.054}^{+0.054}$    & $0.840_{-0.036}^{+0.036}$  \\[1.5mm]
$t_{0}/{\rm Gyr}$       &$13.69_{-0.24}^{+0.25}$      & $13.82_{-0.13}^{+0.13}$     & $13.74_{-0.075}^{+0.074}$    & $13.82_{-0.084}^{+0.085}$    & $13.763_{-0.072}^{+0.071}$  \\[1.5mm]
$z_{\rm re}$            &$10.4_{-1.2}^{+1.2}$       & $10.3_{-1.2}^{+1.2}$        & $10.2_{-1.1}^{+1.1}$         & $10.3_{-1.2}^{+1.2}$         & $10.2_{-1.1}^{+1.1}$  \\[1.5mm]
$h$                     &$0.77_{-0.14}^{+0.15}$       & $0.683_{-0.056}^{+0.054}$   & $0.713_{-0.020}^{+0.020}$    & $0.674_{-0.025}^{+0.025}$    & $0.701_{-0.016}^{+0.016}$  \\[1.5mm]
$D_{\rm V}(z_{\rm m})/{\rm Mpc}$  &$ 1993_{-90}^{+96}$          & $ 2045_{-40}^{+40}$         & $2018_{21}^{+21}$            & $2044_{-19}^{+19}$           & $2030_{-16}^{+16}$  \\[1.5mm]
$f(z_{\rm m})$          &$ 0.755_{-0.025}^{+0.025}$   & $0.754_{-0.022}^{+0.022}$   & $0.760_{-0.016}^{+0.016}$    & $0.748_{-0.019}^{+0.019}$    & $0.764_{-0.015}^{+0.015}$  \\
\hline
\end{tabular}
\label{tab:wde}
\end{table*}

\begin{table*} 
\centering
  \caption{
    The marginalized 68\% allowed regions on the cosmological parameters of the $\Lambda$CDM model extended by allowing for variations on $w_{\rm DE}(a)$ (parametrized according to equation~\ref{eq:wa}),
    obtained using different combinations of the datasets described in Section~\ref{ssec:cmass}
    and \ref{sec:moredata}.}
    \begin{tabular}{@{}lccccc@{}}
    \hline
& \multirow{2}{*}{CMB}  & \multirow{2}{*}{CMB + CMASS} & CMB + CMASS & CMB + CMASS & CMB + CMASS \\
&                       &                           &  +SN     &    +BAO  &   + BAO + SN       \\  
\hline
 $w_0$                  & $-1.12_{-0.51}^{+0.52}$      & $-1.12_{-0.58}^{+0.61}$      & $-1.09_{-0.11}^{+0.11}$      & $-0.95_{-0.27}^{+0.27}$      &$-1.08_{-0.11}^{+0.11}$\\[1.5mm] 
$w_a$                   & $-0.3_{-1.7}^{+1.2}$         & $0.32_{-0.99}^{+0.98}$       & $0.12_{-0.47}^{+0.48}$       & $0.05_{-0.61}^{+0.62}$  &$0.23_{-0.42}^{+0.42}$\\[1.5mm] 
100$\Theta$             & $1.0409_{-0.0016}^{+0.0016}$ & $1.0409_{-0.0016}^{+0.0016}$ & $1.0408_{-0.0016}^{+0.0015}$ & $1.0409_{-0.0016}^{+0.0016}$ &$1.0408_{-0.0016}^{+0.0016}$\\[1.5mm] 
100$\omega_b$           & $2.219_{-0.042}^{+0.042}$    & $2.218_{-0.041}^{+0.042}$    & $2.215_{-0.040}^{+0.040}$    & $2.218_{-0.042}^{+0.00042}$  &$0.0221_{-0.041}^{+0.041}$\\[1.5mm] 
100$\omega_{dm}$        & $11.22_{0.47}^{+0.47}$       & $11.31_{-0.46}^{+0.46}$      & $11.40_{-0.45}^{+0.45}$      & $11.28_{-0.47}^{+0.48}$      &$11.38_{-0.47}^{+0.47}$\\[1.5mm] 
$\tau$                  & $0.0852_{-0.0069}^{+0.0061}$ & $0.0833_{0.0067}^{+0.0062}$  & $0.0823_{-0.0067}^{+0.0058}$ & $0.0833_{-0.0068}^{+0.0061}$ &$0.0825_{-0.0068}^{+0.0060}$\\[1.5mm] 
$n_s$                   & $0.965_{-0.011}^{+0.011}$    & $0.965_{-0.011}^{+0.011}$    & $0.963_{-0.011}^{+0.011}$    & $0.965_{-0.012}^{+0.011}$    &$0.963_{-0.011}^{+0.011}$\\[1.5mm] 
 $\ln(10^{10}A_{\rm s})$ & $3.083_{-0.029}^{+0.030}$    & $3.082_{-0.030}^{+0.030}$    & $3.083_{-0.029}^{+0.029}$    & $3.080_{-0.029}^{+0.029}$    &$3.083_{-0.029}^{+0.030}$\\[1.5mm]
$\Omega_{\rm DE}$       & $0.760_{-0.087}^{+0.081}$    & $0.722_{-0.091}^{+0.081}$    & $0.730_{-0.016}^{+0.016}$    & $0.706_{-0.032}^{+0.032}$    &$0.724_{-0.014}^{+0.014}$\\[1.5mm]
$\Omega_{\rm m}$        & $0.239_{-0.081}^{+0.087}$    & $0.278_{-0.081}^{+0.091}$    & $0.269_{-0.016}^{+0.016}$    & $0.294_{-0.032}^{+0.032}$    &$0.276_{-0.014}^{+0.014}$\\[1.5mm]           
$\sigma_{\rm 8}$        & $0.87_{-0.12}^{+0.12}$       & $0.82_{-0.11}^{+0.11}$       & $0.832_{-0.049}^{+0.049}$    & $0.792_{-0.057}^{+0.057}$    &$0.821_{-0.048}^{+0.048}$\\[1.5mm] 
 $t_{\rm 0}/$Gyr        & $13.64_{-0.22}^{+0.22}$      & $13.79_{-0.16}^{+0.16}$      & $13.763_{-0.091}^{+0.089}$   & $13.827_{-0.086}^{+0.085}$   &$13.80_{-0.083}^{+0.083}$\\[1.5mm] 
  $z_{\rm re}$          & $10.4_{-1.2}^{+1.2}$         & $10.3_{-1.2}^{+1.2}$         & $10.2_{-1.2}^{+1.2}$         & $10.3_{-1.2}^{+1.2}$         &$10.3_{-1.2}^{+1.2}$\\[1.5mm] 
$h$                     & $0.78_{-0.14}^{+0.14}$         & $0.72_{-0.11}^{+0.11}$         & $0.712_{-0.020}^{+0.020}$    & $0.680_{-0.038}^{+0.038}$    &$0.070_{-0.016}^{+0.016}$\\[1.5mm] 
$D_{\rm V}(z_m)/{\rm Mpc}$& $1974_{-83}^{+86}$           & $2040_{-45}^{+47}$           & $2027_{-25}^{+25}$            & $2046_{-20}^{+20}$           &$2038_{-19}^{+19}$\\[1.5mm] 
$f(z_{\rm m})$          & $0.733_{-0.078}^{+0.077}$    & $0.770_{-0.069}^{+0.064}$    & $0.766_{-0.022}^{+0.022}$    & $0.753_{-0.040}^{+0.040}$    &$0.771_{-0.019}^{+0.019}$\\
\hline
\end{tabular}
\label{tab:wa}
\end{table*}

\begin{table*} 
\centering
  \caption{
    The marginalized 68\% allowed regions on the cosmological parameters of the $\Lambda$CDM model extended by allowing for simultaneous variations on $w_{\rm DE}$ (assumed time-independent) and $\Omega_k$,
    obtained using different combinations of the datasets described in Section~\ref{ssec:cmass}
    and \ref{sec:moredata}.}
    \begin{tabular}{@{}lccccc@{}}
    \hline
& \multirow{2}{*}{CMB}  & \multirow{2}{*}{CMB + CMASS} & CMB + CMASS & CMB + CMASS & CMB + CMASS \\
&                       &                           &  +SN     &    +BAO  &   + BAO + SN       \\  
\hline
$\Omega_K$             &  $-0.026_{-0.033}^{+0.028}$   & $-0.0029_{-0.0064}^{+0.0068}$& $-0.0051_{-0.0048}^{+0.0048}$ & $-0.0013_{-0.0061}^{+0.0064}$ &  $-0.0054_{-0.0044}^{+0.0044}$\\[1.5mm]
   $w_{\rm DE}$        &  $-0.91_{-0.47}^{+0.46}$      & $-1.07_{-0.38}^{+0.34}$      & $-1.070_{-0.078}^{+0.079}$    & $-0.946_{-0.16}^{+0.16}$      &  $-1.060_{-0.075}^{+0.075}$\\[1.5mm]
100$\Theta$            &  $1.0410_{-0.0016}^{+0.0016}$ & $1.0412_{-0.0016}^{+0.0016}$ & $1.0411_{-0.0016}^{+ 0.0016}$ & $1.0411_{-0.0015}^{+0.0015}$  &  $1.041_{-0.0016}^{+0.0016}$\\[1.5mm] 
100$\omega_b$          & $2.218_{-0.041}^{+0.041}$    & $2.224_{-0.043}^{+0.043}$    & $2.221_{-0.042}^{+0.042}$     & $2.224_{-0.037}^{+0.038}$     &  $2.220_{-0.041}^{+0.041}$\\[1.5mm] 
100$\omega_{\rm dm}$   & $11.19_{-0.47}^{+0.47}$      & $11.16_{-0.45}^{+0.45}$      & $11.18_{-0.44}^{+0.44}$       & $0.11_{-0.46}^{+0.46}$        &  $11.18_{-0.44}^{+0.44}$\\[1.5mm] 
$\tau$                 & $0.0839_{-0.0071}^{+0.0060}$ & $0.0843_{-0.0066}^{+0.0062}$ & $0.0844_{-0.0068}^{+0.0063}$  & $0.0850_{-0.0060}^{+0.0055}$  &  $0.0850_{-0.0069}^{+0.0059}$\\[1.5mm] 
   $n_s$               & $0.964_{-0.011}^{+0.011}$    & $0.966_{-0.011}^{+0.011}$    & $0.964_{-0.010}^{+0.010}$     & $0.965_{-0.010}^{+0.010}$     &  $0.964_{-0.011}^{+0.011}$\\[1.5mm] 
$\ln(10^{10}A_{\rm s})$& $3.078_{-0.031}^{+0.030}$    & $3.079_{-0.030}^{+0.030}$    & $3.079_{-0.030}^{+0.029}$     & $3.081_{-0.029}^{+0.028}$     &  $3.080_{-0.030}^{+0.030}$\\[1.5mm]
     $\Omega_{\rm DE}$ & $0.62_{-0.18}^{+0.17}$       & $0.725_{-0.064}^{+0.069}$    & $0.733_{-0.017}^{+0.017}$     & $0.707_{-0.027}^{+0.0273}$    &  $0.730_{-0.014}^{+0.014}$\\[1.5mm] 
          $\Omega_m$   & $0.40_{-0.20}^{+0.21}$       & $0.277_{-0.064}^{+0.059}$    & $0.271_{-0.016}^{+0.016}$     & $0.294_{-0.023}^{+0.023}$     &  $0.275_{-0.012}^{+0.012}$\\[1.5mm] 
    $\sigma_{\rm 8}$   & $0.77_{-0.13}^{+0.14}$       & $0.826_{-0.11}^{+0.12}$      & $0.832_{-0.036}^{+0.036}$     & $0.795_{-0.061}^{+0.063}$     &  $0.829_{-0.035}^{+0.035}$\\[1.5mm] 
  $t_{\rm 0}/$Gyr      & $14.7_{-1.1}^{+1.1}$         & $13.90_{-0.23}^{+0.24}$      & $13.97_{-0.23}^{+0.23}$       & $13.88_{0.22}^{+0.21}$        &  $13.99_{-0.20}^{+0.20}$\\[1.5mm] 
        $z_{\rm re}$   & $10.3_{-1.2}^{+1.2}$         & $10.3_{-1.1}^{+1.1}$         & $10.3_{-1.2}^{+1.12}$         & $10.4_{1.1}^{+1.1}$           &  $10.4_{-1.1}^{+1.1}$\\[1.5mm] 
$h$                    & $0.62_{-0.16}^{+0.17}$       & $0.707_{-0.079}^{+0.087}$    & $0.703_{-0.021}^{+0.021}$     & $0.677_{-0.029}^{+0.028}$     &  $0.698_{-0.016}^{+0.016}$\\[1.5mm] 
$D_{\rm V}(z_{\rm m})$ & $2245_{-277}^{+282}$         & $2061_{-40}^{+40}$           & $2054_{-40}^{+40}$            & $2053_{-31}^{+30}$            &  $2061_{-30}^{+30}$\\[1.5mm] 
      $f(z_{\rm m})$   & $0.814_{-0.073}^{+0.073}$    & $0.768_{-0.037}^{+0.041}$    & $0.767_{-0.017}^{+0.017}$     & $0.754_{-0.026}^{+0.026}$     &  $0.768_{-0.015}^{+0.015}$\\ 
\hline
\end{tabular}
\label{tab:wok}
\end{table*}

\begin{table*} 
\centering
  \caption{
    The marginalized 68\% allowed regions on the cosmological parameters of the $\Lambda$CDM model and its extensions, 
    obtained by combining the CMB data with the correlation function of the NGC CMASS sub-sample.}
    \begin{tabular}{@{}lcccccc@{}}
    \hline
& $\Lambda$CDM          & $\Lambda$CDM$+\Omega_k$        & $\Lambda$CDM$+f_{\nu}$         & $\Lambda$CDM$+r$               & $\Lambda$CDM$+w_{\rm DE}$  & $\Lambda$CDM$+w_{\rm DE}(a)$ \\
\hline
$100\Theta$         & $    1.0411_{-0.0015}^{+0.0015}$    & $    1.0411_{-0.0016}^{+0.0015}$ & $    1.0411_{-0.0014}^{+0.0014}$   &  $1.0411_{-0.0015}^{+0.0015}$  & $1.0409_{-0.0016}^{+0.0016}$ & $1.0409_{-0.0016}^{+0.0016}$\\[1.5mm]
$\omega_{\rm dm}$   & $    11.14_{-0.28}^{+0.28}$         & $    11.18_{-0.46}^{+0.46}$      & $11.23_{-0.28}^{+0.28}$       & $11.10_{-0.28}^{+0.28}$   &   $2.217_{-0.042}^{+0.042}$    & $2.217_{-0.041}^{+0.041}$\\[1.5mm]
100\,$\omega_{\rm b}$ &$    2.223_{-0.039}^{+0.039}$       & $    2.223_{-0.040}^{+0.040}$    & $2.224_{   -0.039}^{+0.039}$ &  $2.234_{-0.041}^{+0.041}$  &  $11.32_{-0.46}^{+0.47}$      & $11.29_{-0.45}^{+0.46}$\\[1.5mm]
$\tau $             &  $    0.0850_{-0.0067}^{+0.0059}$   & $    0.0848_{-0.0069}^{+0.0058}$ & $0.0862_{-0.0076}^{+0.0067}$ & $0.0842_{-0.0066}^{+0.0060}$   & $0.0833_{-0.0065}^{+0.0059}$ & $0.0836_{-0.0068}^{+0.0060}$\\[1.5mm]
$n_{\rm s}$         & $    0.9666_{-0.0095}^{+0.0092}$    & $    0.966_{-0.011}^{+0.011}$    & $0.9678_{-0.0095}^{+0.0095}$ & $0.971_{-0.010}^{+0.010}$   &  $0.9636_{-0.0111}^{+0.0111}$ & $0.9641_{0.0111}^{+0.0111}$\\[1.5mm]
$\ln(10^{10}A_{\rm s})$ & $    3.080_{-0.028}^{+0.029}$   & $  3.080_{-0.023}^{+0.030}$     & $3.077_{-0.031}^{+0.031}$     & $3.079_{-0.029}^{+0.029}$   &  $3.082_{-0.029}^{+0.029}$    & $3.082_{-0.029}^{+0.029}$\\[1.5mm]
$\Omega_{k}$        & $                  0             $  & $ -0.0002_{-0.0049}^{+0.0049}$    & $               0      $    & 0   &  $0$                         & $0$\\[1.5mm]
$f_{\nu}$           &  $                  0            $  &                   0          &  $    <0.044$ (95\% CL)          &  $0$  &  $0$                         & $0$\\[1.5mm]
$r$                 & $                  0             $  &                   0          &  $                0      $       & $<0.17$ (95\% CL)   & $0$                         & $0$\\[1.5mm]
$w_{\rm DE}\,(w_0)$ & $                 -1             $  &                  $-1$          &  $               -1      $       &  $-1$  &  $-1.14_{-0.27}^{+0.26}$      & $-1.21_{-0.79}^{+0.61}$\\[1.5mm]
$w_a$               & $                 0              $  &                  $-1$          &  $               -1      $       &  $0$  & $0$                          & $0.14_{-1.0}^{+1.0}$\\[1.5mm]
$\sum m_{\nu}$           &  $                  0            $  &                   0          &  $ <0.52\,{\rm eV}$ (95\% CL)    &  $0$  &  $0$                          & $0$\\[1.5mm]
$\Omega_{\rm DE}$   & $    0.735_{-0.014}^{+0.014}$       & $0.733_{-0.015}^{+0.015}$    &  $    0.723_{-0.017}^{+0.016}$   & $0.738_{-0.014}^{+0.014}$   &  $0.753_{-0.042}^{+0.045}$    & $0.756_{-0.081}^{+0.071}$\\[1.5mm]
$\Omega_{\rm m}$    & $    0.265_{-0.014}^{+0.014}$       & $0.267_{-0.015}^{+0.015}$    &  $    0.276_{-0.016}^{+0.017}$   & $    0.262_{-0.014}^{+0.014}$   &  $0.246_{-0.045}^{+0.042}$    & $0.244_{-0.071}^{+0.081}$\\[1.5mm]
$\sigma_{8}$        &  $    0.813_{-0.018}^{+0.018}$      & $0.814_{-0.025}^{+0.024}$    &  $    0.764_{-0.042}^{+0.040}$   & $0.811_{-0.018}^{+0.018}$   &  $0.861_{-0.09}^{+0.10}$      & $0.87_{-0.11}^{+0.11}$\\[1.5mm]
$t_{0}/{\rm Gyr}$   &  $   13.727_{-0.071}^{+0.072}$      & $13.74_{-0.24}^{+0.23}$      &  $   13.818_{-0.098}^{+0.10}$    & $13.713_{-0.075}^{+0.074}$   & $13.69_{-0.13}^{+0.13}$      & $13.69_{-0.14}^{+0.14}$\\[1.5mm]
$z_{\rm re}$        &  $   10.4_{-1.1}^{+1.1}$            & $10.4_{-1.1}^{+1.1}$         &  $   10.5_{-1.2}^{+1.3}$         & $   10.3_{-1.1}^{+1.2}$   &  $10.3_{-1.1}^{+1.1}$         & $10.3_{-1.1}^{+1.2}$\\[1.5mm]
$h$                 &   $   0.711_{-0.012}^{+0.012}$ & $0.708_{-0.017}^{+0.017}$    &  $   0.698_{-0.015}^{+0.015}$    & $0.713_{-0.013}^{+0.013}$   &  $0.751_{-0.073}^{+0.078}$    & $0.77_{-0.12}^{+0.12}$\\[1.5mm]
 $D_{\rm V}(z_{\rm m})/{\rm Mpc}$  &  $ 2005_{-20}^{+20}$ & $2010_{-40}^{+40}$           &  $ 2030_{-27}^{+28}$             & $ 2001_{-22}^{+21}$   &  $1997_{-37}^{+37}$           & $1996_{-41}^{+43}$\\[1.5mm]
$f(z_{\rm m})$          &  $0.754_{-0.012}^{+0.012}$      & $0.745_{-0.013}^{+0.013}$    &  $    0.753_{-0.014}^{+0.014}$   &  $0.740_{-0.012}^{+0.012}$  & $0.757_{-0.027}^{+0.028}$    & $0.760_{-0.071}^{+0.064}$\\ 
\hline
\end{tabular}
\label{tab:ngc}
\end{table*}

\end{document}